# Ionisation processes and laser induced periodic surface structures in dielectrics with mid-infrared femtosecond laser pulses

George D. Tsibidis[1,*] and Emmanuel Stratakis[1,2]

[1] Institute of Electronic Structure and Laser (IESL), Foundation for Research and Technology (FORTH), N. Plastira 100, Vassilika Vouton, 70013, Heraklion, Crete, Greece
[2] Department of Physics, University of Crete, 71003 Heraklion, Greece

*tsibidis@iesl.forth.gr

## ABSTRACT

Irradiation of solids with ultrashort pulses and laser processing in the mid-Infrared (mid-IR) spectral region is a yet predominantly unexplored field with a large potential for a wide range of applications. In this work, laser driven physical phenomena associated with processes following irradiation of fused silica ($SiO_2$) with ultrashort laser pulses in the mid-IR region are investigated in detail. A multiscale modelling approach is performed that correlates conditions for formation of perpendicular or parallel to the laser polarisation low spatial frequency periodic surface structures for low and high intensity mid-IR pulses (not previously explored in dielectrics at those wavelengths), respectively. Results demonstrate a remarkable domination of tunneling effects in the photoionisation rate and a strong influence of impact ionisation for long laser wavelengths. The methodology presented in this work is aimed to shed light on the fundamental mechanisms in a previously unexplored spectral area and allow a systematic novel surface engineering with strong mid-IR fields for advanced industrial laser applications.

## Introduction

The employment of ultra-short pulsed laser sources for material processing has received considerable attention over the past decades due to the important technological applications, particularly in industry and medicine [1-7]. There is a variety of surface structures generated by laser pulses and more specifically, the so-called laser-induced periodic surface structures (LIPSS) on solids that have been studied extensively [8-19] and are related to those applications. A range of LIPSS types have been produced based on the material and the laser parameters [6,9,16,20,21]. According to the morphological features of the induced surface structures such as their periodicity and orientation, LIPSS can be classified in: (a) High Spatial Frequency LIPSS (HSFL) [16,22], (b) Low Spatial Frequency LIPSS (LSFL) [10,23], (c) Grooves [18,21], (d) Spikes [18] and (e) complex ones [20,21,24]. The LIPSS fabrication technique as well as the associated laser driven physical phenomena have been the topic of an extensive investigation. This is due to the fact that the technique constitutes a precise, single-step and scalable method to fabricate highly ordered, multi-directional and complex surface structures that mimic the unique morphological features of certain species found in nature, an approach which is usually coined as biomimetics. A thorough knowledge of the fundamental mechanisms that lead to the LIPSS formation provides the possibility of generating numerous and unique surface biomimetic structures [2,6,20,25-27] for a range of applications, including microfluidics [1,28], tribology [29-31], tissue engineering [5,28] and advanced optics [6,32].

On the other hand, a key characteristic of the investigation of the underlying physical processes that account for LIPSS formation is that it has been centred on laser pulses in a spectral region between the visible and near-infrared frequencies ($\lambda_L < 1.5$ μm) [21,33]. Nevertheless, there is recently an increasing interest in the nano/micro sized patterning of materials with longer wavelengths. The motivation to explore the response of the irradiated materials and relevant surface effects in the mid-IR region originates from the challenging opportunities in photonics for mid-IR radiation [34-36]. Mid-IR has proven to constitute a spectral region that provides numerous possibilities for both fundamental and applied research [37]. For example, irradiation with intense mid-IR lead to novel phenomena like photon acceleration in metasurfaces to generation of attosecond pulses, and demonstration of megafilamentation in atmosphere, which may open doors to many exciting new applications [38]. Similarly, mid-IR-assisted damage can be used for the sub-surface micromachining of multi-layer materials, such as the inscription of silicon waveguides operating in the telecommunications band [39]. On the other hand, an abundance of molecules have impressive absorption features in the mid-IR region which makes them highly useful for applications in biomolecular sensing, explosives detection, biomedical applications, and environmental sensing [40].

Despite the above exciting possibilities and the advances in the investigation of behaviour of irradiated material with mid-IR pulses [37-39,41-44], there are still many crucial questions that have yet to be addressed. A fundamental question is, however, whether the underlying physics that characterises laser-matter interaction for mid-IR differs from that at lower spectral regions [45-48]. In a previous report, results showed that silicon is transparent in this region while it absorbs strongly during the pulse duration [43]. Furthermore, a systematic analysis of the role of excitation levels for conditions directly related to formation of sub-wavelength LIPSS through the investigation of surface plasmon excitation-based mechanisms revealed significantly different effects, such as surface plasmons (SP) with smaller confinement, longer lifetime and larger decay lengths for mid-IR pulses compared to irradiation with lower wavelengths [43]. These results demonstrate the need to describe in more detail the ultrafast dynamics following excitation with mid-IR sources as they can potentially influence the morphological changes on the material [43]. Another important characteristic of irradiation of solids with mid-IR sources is that as electron cycle average energy scales as $\lambda_L^2$ ($\lambda_L$ stands for the laser wavelength), excited electrons are expected to gain enough energy to modify damage mechanisms at long wavelengths. Therefore, a detailed description of the physical processes that characterise interaction of laser sources with solids is crucial for efficient laser-based machining in the mid-IR region ($\lambda_L$ >2 μm).

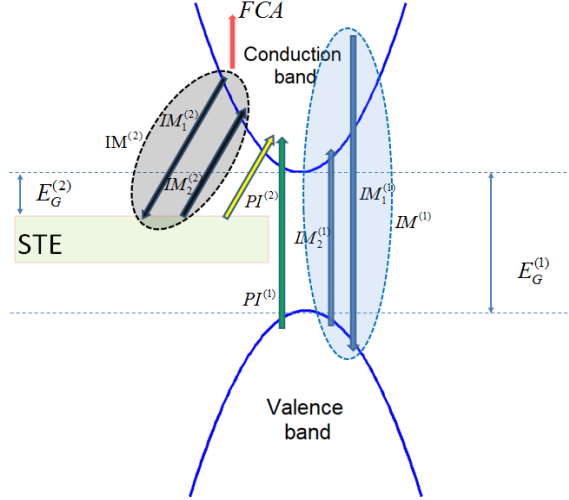

**Figure 1.** Excitation processes in fused silica: *Green* and *yellow* arrows indicate photoionisation mechanism ($PI^{(1)}$ and $PI^{(2)}$) from VB and self-trapped exciton (STE) levels, respectively. The position of the STE 'box' simply illustrates graphically that the STE band lies between VB and CB. *Blue* region corresponds to the impact ionisation process $IM^{(1)}$ from CB to VB [*blue* down arrow indicates collision of highly energetic electrons in CB with electrons in VB ($IM_1^{(1)}$) while *blue* up arrow indicates transfer of the latter into CB ($IM_2^{(1)}$)]. Similarly, *gray* region corresponds to the impact ionisation ($IM^{(2)}$) from CB to STE [*black* down arrow indicates collision of highly energetic electrons in CB with electrons in STE ($IM_1^{(2)}$) while *black* up arrow indicates transfer of the latter into the CB) ]. $E_G^{(1)}$ (or $E_G^{(2)}$) stand for the band gaps between top of VB (or STE) and bottom of CB. *Red* arrow corresponds to inverse bremsstrahlung (free carrier absorption, FCA).

One type of material which is of paramount importance for the design of optics in high power laser systems, is fused silica. While an extensive research has been conducted towards elucidating laser-induced damage in fused silica through the investigation of processes related to irradiating $SiO_2$ with IR pulses [6,11,12,17,22,48-54], little is known about the effects of electron excitation with pulses of longer wavelengths [55]. To account for the influence of mid-IR photons on the physical processes related to excitation conditions as well as possible surface modification-related mechanisms, a number of critical factors should be considered: (i) the energy absorption and excitation at larger $\lambda_L$, (ii) the significance of nonlinear processes such as Kerr effect, multi-photon absorption, tunneling effect and impact ionisation, (iii) the role of the wavelength value in the modulation of the optical parameters, and (iv) the conditions that lead to surface patterning. In regard to the surface modification, LSFL structures with orientation parallel to that of the electric field of the incident beam (termed here as LSFL∥) have been predicted and observed for moderate intensities at lower wavelengths [11,22,48,53]; these predictions were based on Sipe's theory [19] which proposes that surface patterning results from inhomogenous energy absorption from the solid due to surface roughness. Similarly, LSFL structures with orientation perpendicular to that of the electric field of the incident beam (termed here as LSFL⊥) have also been observed in dielectrics, when intense laser beams can potentially lead to excitation levels for which a transition of the dielectric material to a 'metallic' state can be achieved [22,53,56]. Theoretical simulations which are validated with experimental results support the proposed scenario that the interference of the incident light with the far-field of rough non-metallic or metallic surfaces accounts for the formation of both LSFL∥ and LSFL⊥ [22,53]. On the other hand, for metallic surfaces, SP-related processes have been, also, proposed to successfully describe LSFL⊥ structures with periodicities ~ $\lambda_L$ [10,57]. Therefore, an interesting question that needs to be explored is whether conditions for the formation LSFL∥ and LSFL⊥ can, also, be achieved following irradiation of $SiO_2$ with mid-IR sources.

In this paper, we report on the excitation of $SiO_2$ following irradiation with ultrashort (femtosecond) pulses at various wavelengths in the mid-IR spectral region (between $\lambda_L$=2 μm and $\lambda_L$=4 μm). The photoionisation rates are computed and the role of tunneling effect is highlighted. The impact of Kerr effect and nonlinearities in the refractive index $n$ of the material as well as the resulting electron densities is analysed. The ultrafast dynamics of the excited carriers is described in the presence of metastable self-trapped exciton (STE) states [49]. Formation of LSFL∥ and LSFL⊥ and their periods are investigated for a range of electron density values at various laser intensities based on Sipe [8,19] and SP [58] theories, respectively. Finally, a Two Temperature Model (TTM) [59] coupled with a hydrodynamical module [10,11] is used to describe energy transfer between the electron and the lattice systems and surface modification processes assuming that surface patterning requires a phase transformation and melting of the heated material.

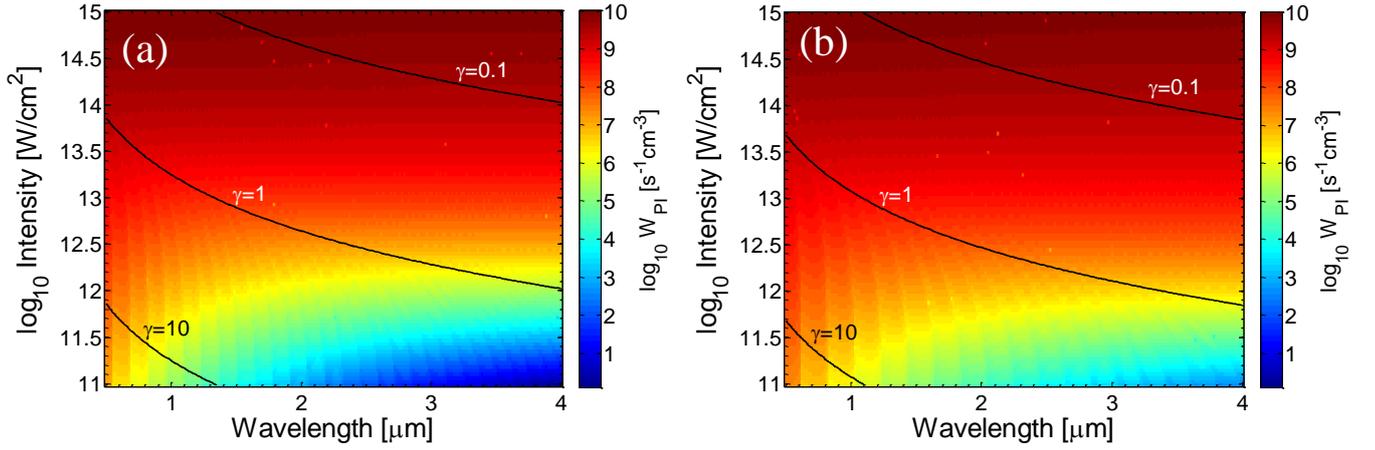

**Figure 2.** Photoionisation rates at various wavelengths and laser (peak) intensities assuming excitation of electrons from VB (a) and STE (b). The *black* lines define regions in which the Keldysh parameter $\gamma$ attains values $\gamma=0.1, 1, 10$.

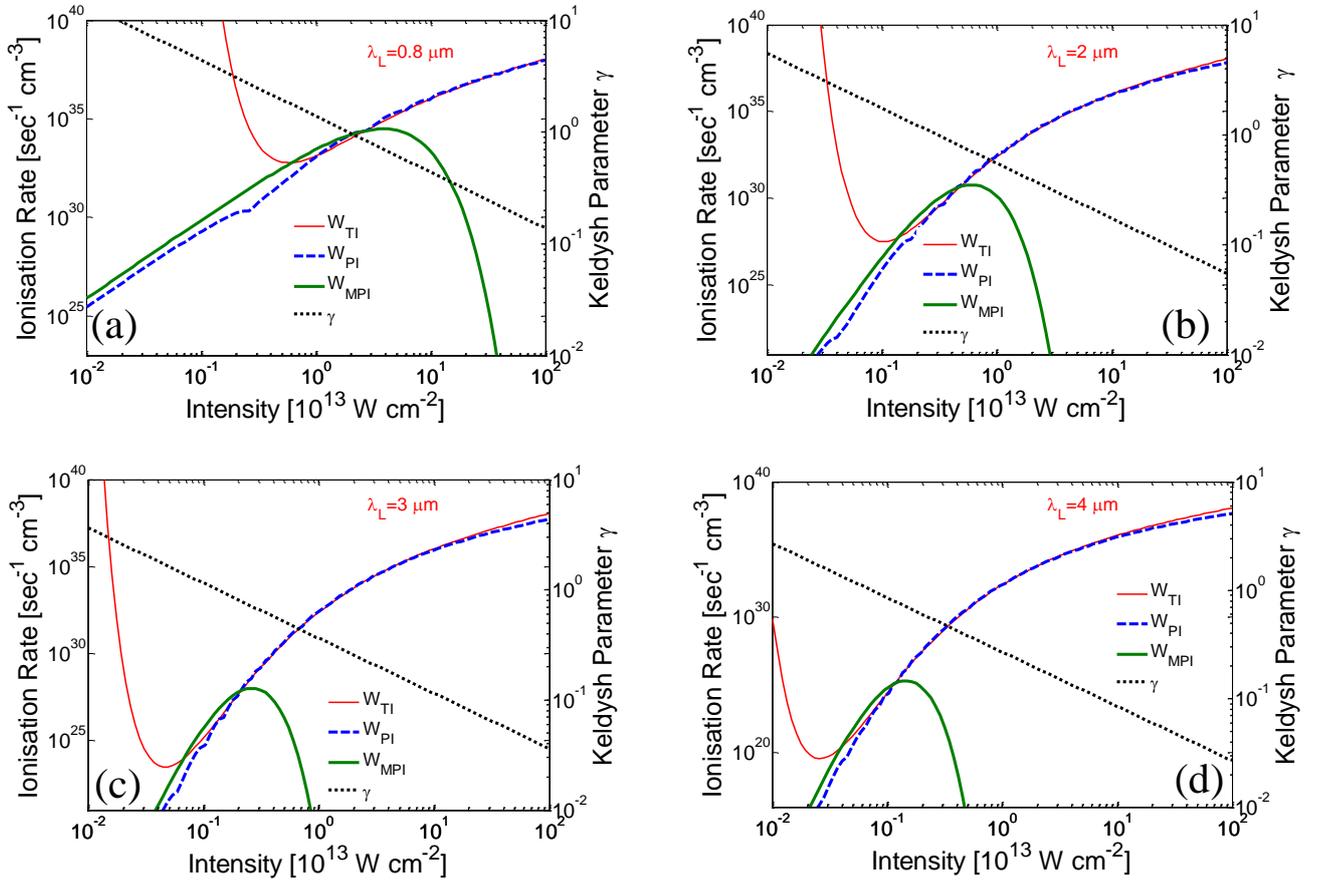

**Figure 3.** Photoionisation rates as a function of the laser (peak) intensity. Intensities at which multiphoton, tunneling and combined photoionisation dominate depending of the Keldysh number are illustrated. Results are shown for $\lambda_L=0.8$ μm, 2 μm, 3 μm, 4 μm.

## Results and Discussion

In this Section, the theoretical model that describes the electron excitation following irradiation of $SiO_2$ with mid-IR pulses is presented. A more detailed presentation of the mathematical model is provided in the Supplementary material. Furthermore, the fundamental mechanisms of surface modification and the evaluation of the surface pattern periodicities are described through a multiscale theoretical model.

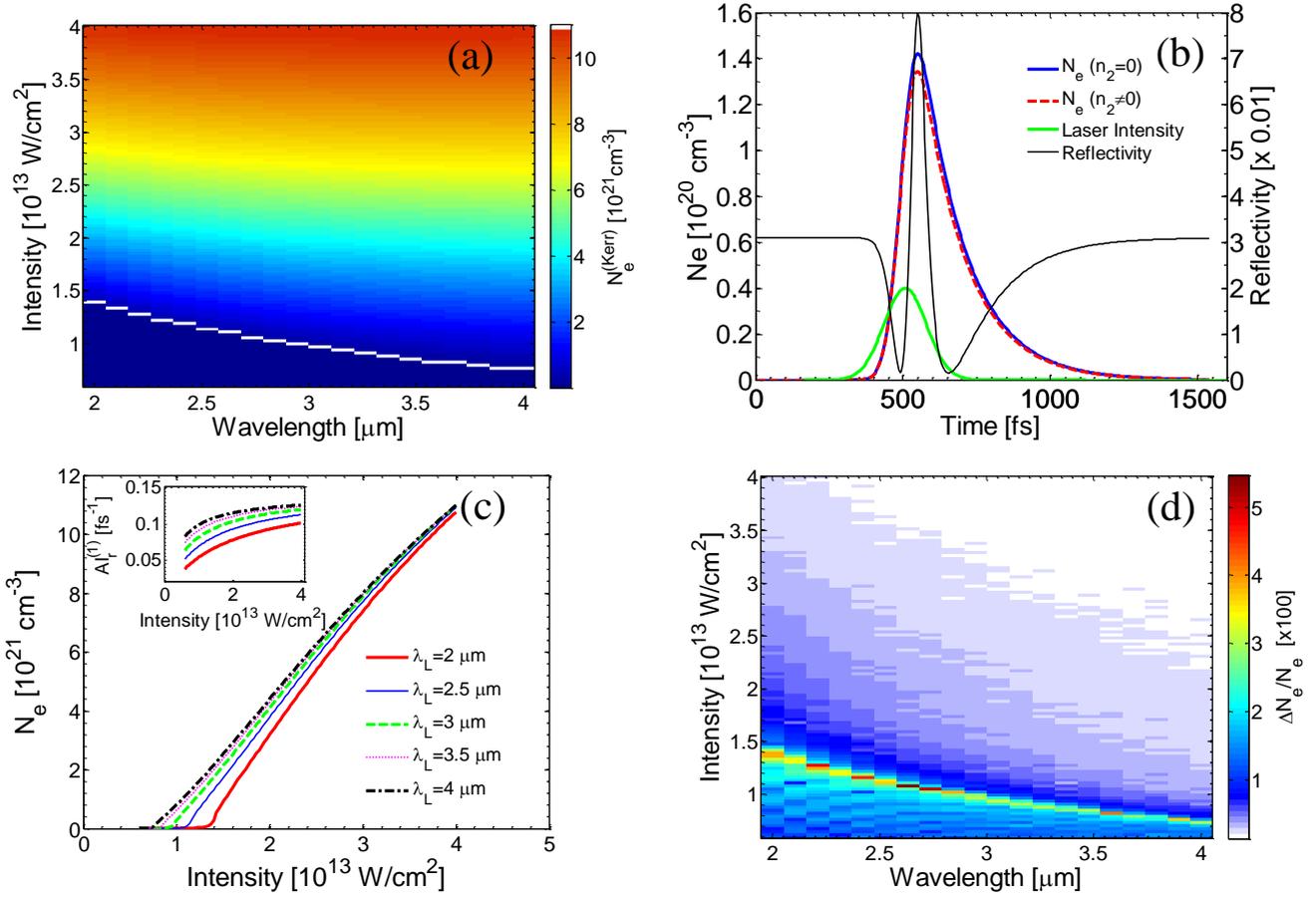

**Figure 4.** (a) MVED at various wavelengths and laser (peak) intensities (*white* line defines OBT). (b) Electron density evolution for $\lambda_L$=2.6 μm and $I$=1.26×10$^{13}$ W/cm$^2$ with and without Kerr effect. Laser pulse is shown in arbitrary units (*green* line). (c) MVED dependence on laser intensity (Results are shown for $\lambda_L$=2 μm, 2.5 μm, 3 μm, 3.5 m and 4 μm). Inset illustrates the influence of impact ionisation. (d) Percentage change of MVED if Kerr effect is considered ($\Delta N_e = \left(N_e - N_e^{(Kerr)}\right)$). ($\tau_p$=170 fs).

**Photoionisation processes.** In the absence of defects, the absorption of light in dielectrics and photoionisation (PI) is a nonlinear process because a single photon does not have enough energy to excite electrons from the valence (VB) to the conduction band (CB). In the presence of intense laser fields, both multiphoton (MPI) and tunneling ionisation (TI) can occur in different regimes depending on the laser intensity values [50,60]. A detailed description of the ionisation processes is illustrated in Fig.1. The free electrons in CB generated through PI operate as 'seeds' for a dominant excitation process, the impact ionisation [50]. The latter can occur once the conduction band electrons have sufficiently high energy; a part of their energy may be transferred to bound electrons in VB by collision (process $IM_1^{(1)}$ in Fig.1) resulting in free electrons at the bottom of CB (process $IM_2^{(1)}$ in Fig.1). These free electrons will subsequently experience the same process as discussed above, and produce more electrons in CB by exciting electrons from VB. On the other hand, relaxation processes lead to the production of metastable STE states [61]. In Fig.1, STE states are illustrated as centres/defects situated in an energy region $E_G^{(2)}$=6 eV [49] below the minimum of CB (smaller than the band gap of fused silica, $E_G^{(1)}$=9 eV). Trapping time of electrons in STE states is taken to be $\tau_{tr}$~150 fs [62]. If the laser pulse duration is *longer* than $\tau_{tr}$ and given that STE decay time is of the range of some hundreds of picoseconds [63,64], excitation through photoionisation ($PI^{(2)}$) and impact ionisation ($IM^{(2)}$ through sequential processes $IM_1^{(2)}$ and $IM_2^{(2)}$ similar to those for $IM^{(1)}$ in Fig.1) of STE states are possible [49,61]. Furthermore, electrons in CB can further absorb energy from the laser photons through a linear process, the free carrier absorption (denoted by FCA label in Fig.1) [50], and move to higher states in CB.

As stated above, PI depends on the laser intensity, however, MPI and TI are efficient in different regimes of the intensity spectrum. In Fig.2, the photoionisation rate $W_{PI}$ is computed for (peak) intensities $I$ in the range between 10$^{11}$ W/cm$^2$ and 10$^{15}$ W/cm$^2$ and for laser wavelengths $\lambda_L$=0.5-4 μm. Fig.2a,b describe photoionisation both of VB and STE electrons. The Keldysh parameter, $\gamma \sim 1/(\sqrt{I}\lambda_L)$, indicates the regime in which each of the photoionisation components, MPI and TI dominates [60]. It is evident that at longer wavelengths and large intensities, TI dominates ($\gamma \ll 1$) while at shorter wavelengths and small intensities, MPI is the main contributor to PI ($\gamma \gg 1$). By contrast, an intermediate regime in which both TI and MPI coexist is illustrated in a region between $\gamma=0.1$ and $\gamma=10$. Previous studies demonstrated that the Keldysh approach to distinguish between the extremes of two different descriptions of ionization is valid in the mid-IR spectral region [42]. In Fig.2, it is shown that pronounced ridges occur at low intensities (Fig.2) for various wavelengths which are justified by the different number of photon energies required for MPI to ionise the material. On the other hand, the wavelength dependence disappears as $\gamma$ moves to regimes where TI dominates which occurs also at smaller wavelengths [41,65]. The approximating values for TI and MPI as a function of the laser intensity at various wavelengths are illustrated in Fig.3. To evaluate the regime where TI or MPI become more efficient in the PI process, the Keldysh parameter is also displayed. Results for three wavelengths in the mid-IR spectral region ($\lambda_L=2$ μm, 3 μm, 4 μm) are compared with calculations for $\lambda_L=800$ nm. It is evident that for large intensities, the Keldysh approximation for MPI provides an underestimation of PI rates while for lower values of $I$ an overestimated TI is revealed. Interestingly, as the laser wavelength increases, the threshold value of $\gamma$ at which MPI is less efficient than TI decreases. Furthermore, while at $\lambda_L=800$ nm TI appears to become a dominant ionisation process at $\gamma \sim 1$, this behaviour occurs at even smaller values of $\gamma$ (<0.1) at longer wavelengths (Fig.3c,d).

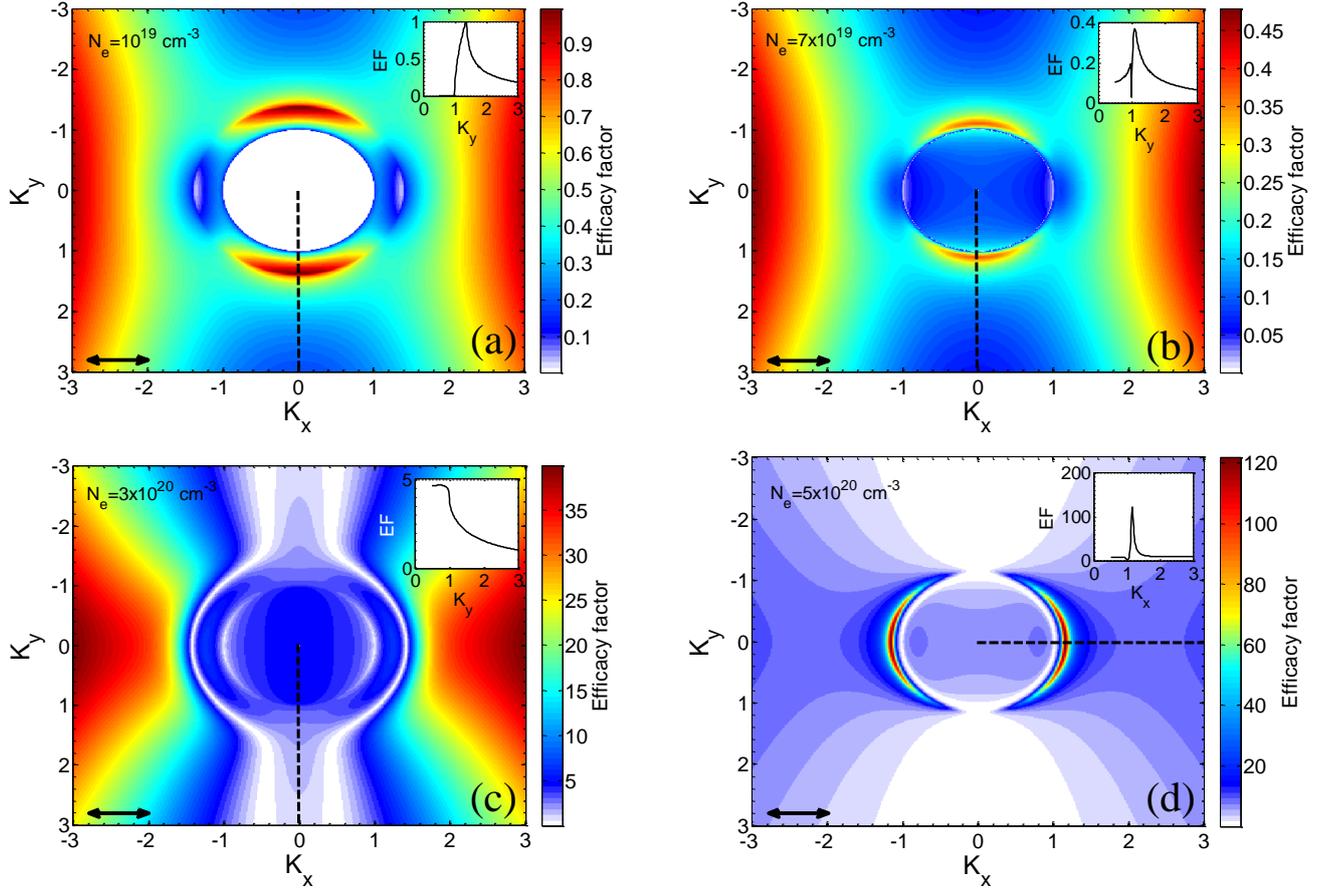

**Figure 5.** Efficacy factor maps for (a) $N_e=10^{19}$ cm$^{-3}$, (b) $N_e=7\times10^{19}$ cm$^{-3}$, (c) $N_e=3\times10^{20}$ cm$^{-3}$ and (d) $N_e=5\times10^{20}$ cm$^{-3}$ ($\lambda_L=2.6$ μm). Inset shows maps across the *black* dashed lines. *Black* double-ended arrows indicate polarisation direction.

**Electron excitation.** To describe the free electron dynamics following excitation with ultrashort pulsed lasers, two single rate equations are used to account for the temporal evolution of the excited electrons from both VB and STE. Based on the scheme of the underlying physical mechanisms for photoionisation described in Fig.1 and assuming laser pulse durations longer than $\tau_r$, excitation of STE to the CB is also considered. It is noted that the proposed model ignores the fact that only electrons in the CB with a sufficiently high kinetic energy are capable to participate in impact ionisation processes and enable electrons in the VB to overcome the ionisation potential. Revised models to remove this overestimation of the induced impact ionisation (with [11] or without [48,51] the inclusion of STE states) has been presented in previous reports in which multiple rate equations were introduced [51]. Nevertheless, the model that is used in this work is based on the following simplified set of two-rate equations (TRE) which have also been used in previous works at lower wavelengths [49,52,62,66]

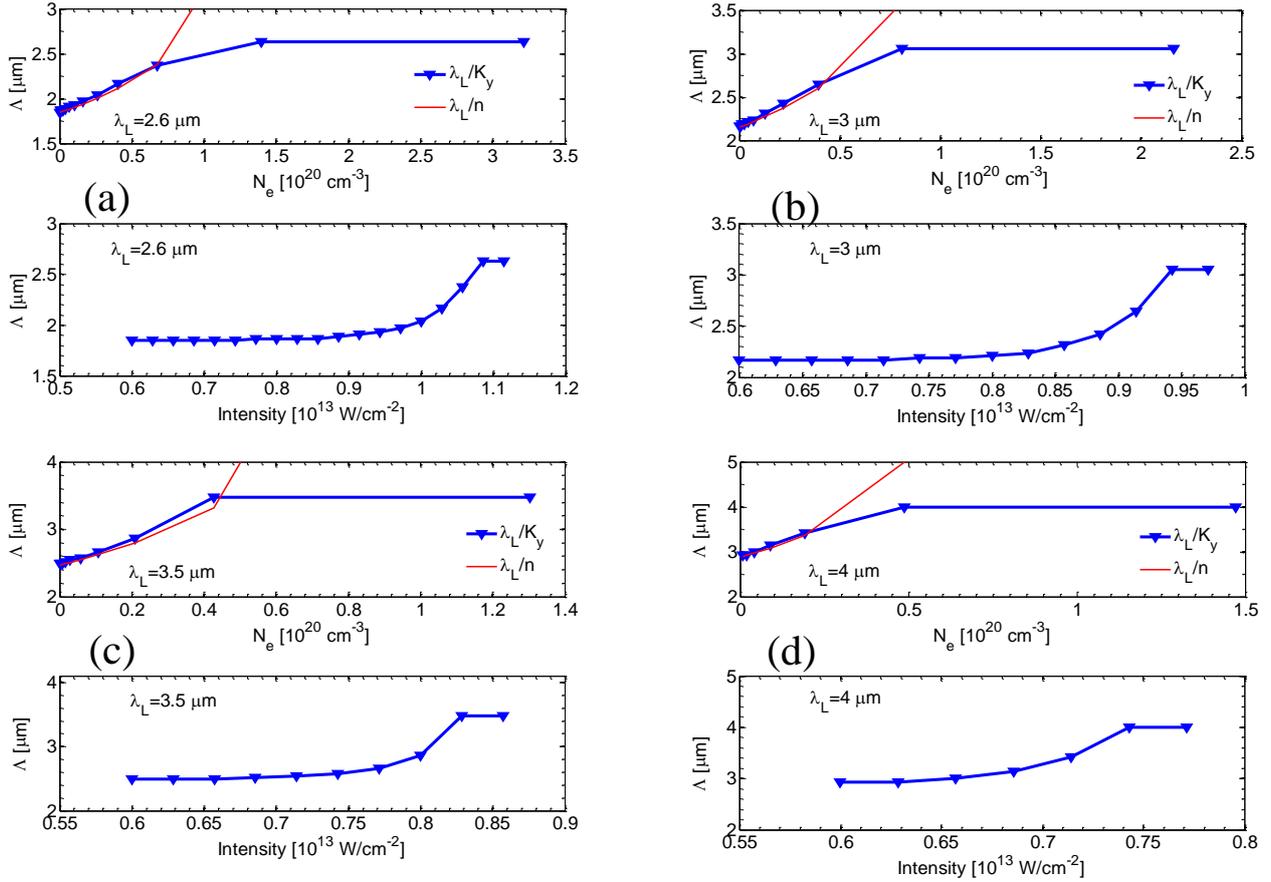

**Figure 6.** Calculated periodicities of parallel subwavelength LIPSS as a function of electron densities and laser (peak) intensities ($\lambda_L$=2.6 μm, 3 μm, 3.5 μm, and 4 μm) ($\tau_p$=170 fs).

$$\begin{aligned}
\frac{dN_e}{dt} &= \frac{N_v - N_e}{N_v}\left(W_{PI}^{(1)} + N_e AI_r^{(1)}\right) + \frac{N_{STE}}{N_v}\left(W_{PI}^{(2)} + N_e AI_r^{(2)}\right) - \frac{N_e}{\tau_{tr}} \\
\frac{dN_{STE}}{dt} &= \frac{N_e}{\tau_{tr}} - \frac{N_{STE}}{N_v}\left(W_{PI}^{(2)} + N_e AI_r^{(1)}\right)
\end{aligned} \quad (1)$$

where $AI_r^{(1)}$ and $AI_r^{(2)}$ are the (usually termed [50,67]) avalache ionisation rate coefficients for the impact ionisation $IM^{(1)}$ and $IM^{(2)}$ processes, $N_V$=2.2×10$^{22}$ cm$^{-3}$ [49] corresponds to the atomic density of the unperturbed material while $N_e$ and $N_{STE}$ denote the free and STE electron densities, respectively. Finally, the photoionisation rates $W_{PI}^{(1)}$ and $W_{PI}^{(2)}$ include both multiphoton and tunneling ionisation processes; as noted in the previous paragraphs, each one of the multiphoton and tunneling ionisation processes become more efficient in different regimes based on the value of $\gamma$ [49,60]. Maximum values of the electron densities (MVED) as a function of the laser wavelength and intensity are illustrated in Fig.4a. It is noted that results for MVED include contribution from nonlinearities in the refractive index $n$ due to Kerr effect (denoted with $N_e^{(Kerr)}$); the influence of Kerr effect is discussed in the next section. In this study, simulations were performed for laser pulse duration equal to $\tau_p$=170 fs at different peak intensities ($I = \frac{2\sqrt{\ln 2}J}{\sqrt{\pi}\tau_p}$ [43]) where $J$ stands for the energy fluence. In the Supplementary Material, simulation results for MVED have been obtained for different values of the pulse duration and fluence to highlight the influence of STE states for single shot irradiation. Similar investigation can be pursued for multi pulse simulated experiments.

The temporal variation of reflectivity for the irradiated material (Fig.4b) resembles the theoretical calculations and experimental results at shorter wavelengths [53,68]. The pulse is switched on at $t$=0 and reaches the peak at $t$=$3\tau_p$ (*green* line in Fig.4b). During the pulse, the reflectivity initially drops while it, gradually, starts to increase when material undergoes a transition to a 'metallic' state; then, the reflectivity reaches a peak value near the end of the pulse before it starts falling again.

An important outcome of the simulations is the dependence of the MVED on the laser intensities at different $\lambda_L$, shown in Fig.4c. It can be observed that the maximum values of the produced electrons for the same laser intensity rises at increasing wavelength. This is a result that has been observed experimentally in silicon [67] at lower wavelengths while it has also been predicted for irradiation of silicon with mid-IR laser pulses [43]. The monotonicity is attributed to the increase of the strength of impact ionisation rate at increasing wavelength which is confirmed by the simulations (inset in Fig.4c). In

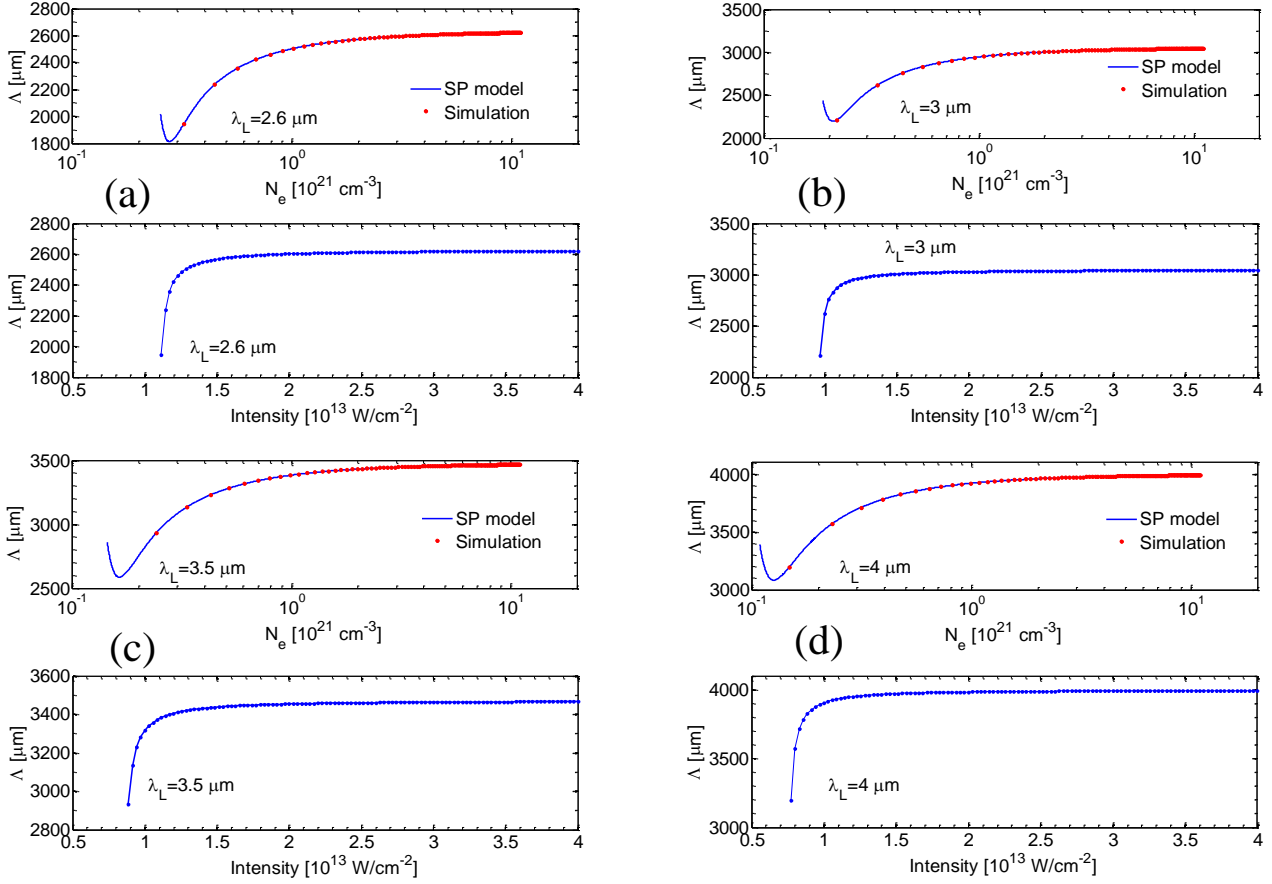

**Figure 7.** Calculated surface plasmon wavelengths as a function of electron densities and laser (peak) intensities ($\lambda_L$=2.6 μm, 3 μm, 3.5 μm, and 4 μm) ($\tau_p$=170 fs).

Fig.4c, the impact ionisation rate illustrated for excitations from VB to CB (denoted by $AI_r^{(1)}$). Similar monotonicity occurs for impact related excitation processes for the STE states.

**Influence of Kerr effect.** To account for the Kerr effect, an intensity dependent variation of the dielectric constant $\Delta\varepsilon = 2n_0 n_2 I + (n_2 I)^2$ is considered where $n_0$ is the refractive index of the unexcited material and $n_2$ is the Kerr coefficient which is related to the Kerr effect (see Supplementary Material). While extensive measurements of the nonlinear refractive index $n_2$ exist for various types of materials in the visible and near-infrared (IR) wavelengths, few such measurements have been performed in the longer, spectroscopically important mid-IR regime. Experimental measurements for fused silica found in recent studies for $n_2$ (equal to (1.94±0.19)×10$^{-16}$ cm$^2$/W and (2.065±0.23)×10$^{-16}$ cm$^2$/W for $\lambda_L$=2.3 μm and 3.5 μm, respectively [69]), indicate a practically constant value for the Kerr coefficient. In principle, $n_2$ for fused silica in the mid-IR spectral region appears to be 100 times smaller than the experimentally measured value for silicon [43]. The *white* line in Fig.4a shows the intensity thresholds (i.e. optical breakdown threshold OBT) for which the simulated $N_e$ exceeds the critical value $N_e^{cr}$ (i.e. $N_e^{cr} \equiv 4\pi^2 c^2 m_e \varepsilon_0 / (\lambda_L^2 e^2)$; $N_e^{cr}$ corresponds to the free electron density at which the plasma oscillation frequency is equal to the laser frequency, where $c$ is the speed of light, $m_e$ stands for the electron mass, $e$ is the electron charge and $\varepsilon_0$ is the permittivity of vacuum). $N_e^{cr}$ is, often, associated with the induced damage on the material following exposure to intense heating, however, as it will be explained later in the text, a thermal criterion will be used in this work to describe damage [62]. To evaluate the influence of Kerr effect in the electron excitation, a series of simulations were performed to highlight the contribution of the Kerr effect at different intensities and wavelengths. In Fig.4b, simulations for $I$=1.26×10$^{13}$ W/cm$^2$ and $\lambda_L$=2.6 μm, show that the maximum electron density lies higher than the MVED if Kerr effect is assumed. Similar conclusions follow at different wavelengths and intensities (Fig.4d). Moreover, if Kerr effect is taken into account, the refractive index which is also a measure of reflectivity, increases as the intensity becomes higher due to the nonlinear part in the expression for $n$ (for small intensity values). As a result, an enhanced reflectivity is produced which leads to a lower energy absorption from the material and therefore a lower excited electron density $N_e^{(Kerr)}$ (Fig.4b).

Relevant simulations for MVED at wavelengths between $\lambda_L$=2 μm and 4 μm and intensities up to $I$=4×10$^{13}$ W/cm$^2$ are illustrated in Fig.4d where a maximum 5% increase of the MVED is calculated. Interestingly, at a given wavelength, as the intensity increases there is an initially increase of the discrepancy which reaches the maximum value at around the intensity that yields MVED = $N_e^{cr}$, before it starts falling at larger values of $I$ (Fig.4d). The rising trend can be attributed to the calculated reflectivity and the resulting absorbed energy; at low intensities, reflectivity is small (Fig.4b) that gives rise to larger differences, if Kerr effect is considered. By contrast, at higher intensities, the Kerr polarizability of the medium rapidly ceases to exist (the band structure is being progressively destroyed) due to ongoing ionisation; thus, the response of the medium is strongly dominated by the behaviour of the free electron-hole plasma which is formed during the pulse. This is followed by an increase of the reflectivity when the free electron density becomes higher than the critical density while it leads to a gradual drop of the electron density. The fact that the intensity values for which the difference $(N_e - N_e^{(Kerr)})/N_e$ becomes maximum appear to coincide with OBT (Fig.4a,4d) requires more investigation. Although, the agreement does not appear to be coincidental, a deeper exploration of the role and interplay of various excitation processes should further analysed. Nevertheless, it is evident that a direct comparison with Kerr effect is not convincing as the expression of $N_e^{cr}$ is independent of the refractive index.

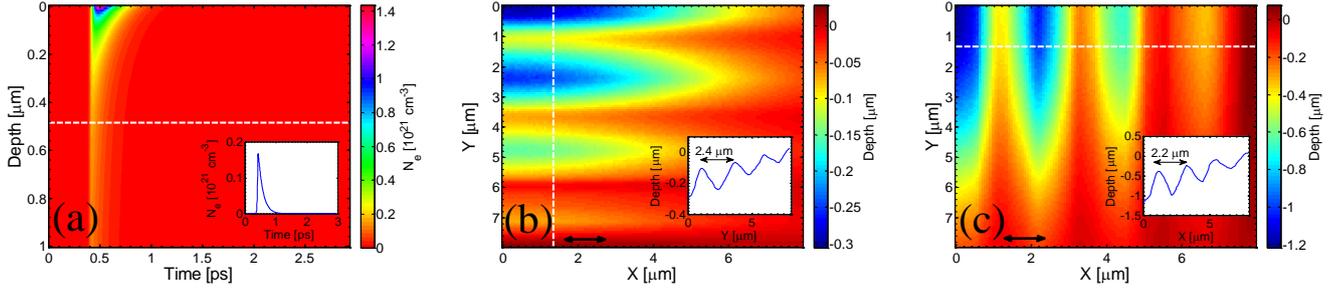

**Figure 8.** (a) Evolution of the spatial distribution of $N_e$ along $z$-axis at position where energy deposition is higher for $NP$=1 (for $I$=1.4×10$^{13}$ W/cm$^2$), (dashed *white* line corresponds to the maximum depth for which OBT occurs while inset shows the temporal profile of $N_e$ along the dashed line). (b) Top view of Parallel and (c) Perpendicular to laser polarisation LSFL for laser (peak) intensities $I$=1.06×10$^{13}$ W/cm$^2$ and $I$=1.4×10$^{13}$ W/cm$^2$, for $NP$=3 and $NP$=10, respectively. *Black* double-ended arrows indicate polarisation direction. (insets in (b) and (c) show the depth profiles along the dashed lines) ($\tau_p$=170 fs). Figures (b) and (c) illustrate a quadrant of the affected zone (top view) ($\lambda_L$=2.6 μm).

**LIPSS formation.** One of the predominant advantages of elucidating the interrelation of the processes taking place in $SiO_2$ following irradiation with femtosecond laser pulses is that it can open unique ways for optimising and controlling patterning methodologies. It is known that energy release into the lattice depends to a large extent on the capability of the irradiated material to transfer electronic excitations into vibrational modes. The aforementioned model (Eq.1) is aimed to enable determination of electron excitation levels and efficient description of electron dynamics which are both correlated with induced morphological changes on the surface or volume of the irradiated material. As explained in the introductory section of this work, LIPSS is one important type of surface patterns which is of paramount importance to a wide range of applications. To describe the formation of LIPSS, the electron density values and surface corrugation features are used to determine the orientation as well as the periodicity of LIPSS. In a similar fashion, as the procedure followed for irradiation with IR pulses [53], a distinction is made about the conditions for formation of LSFL∥ and LSFL⊥. Currently, the most widely accepted theory of LSFL is based on the interference of the incident laser beam with some form of a surface-scattered electromagnetic wave (SP waves could also be included if appropriate conditions are satisfied [58]). To correlate $N_e$ with possible formation of laser-induced surface structures, the inhomogeneous energy deposition into the irradiated material is computed through the calculation of the product $\eta(\vec{K}, \vec{k_L}) \times |b(\vec{K})|$ as described in the model of Sipe [19]. In the above expression, $\eta$ represents the efficacy with which a surface roughness at the wave vector $\vec{K}$ (i.e., normalised wavevector $|\vec{K}|$ =$\lambda_L/\Lambda$, where $\Lambda$ stands for the predicted structural periodicity) induces inhomogeneous radiation absorption, $\vec{k_t}$ is the component of the wave vector of the incident laser beam on the material's surface plane and $b$ represents a measure of the amplitude of the surface roughness at $\vec{K}$. In principle, surface roughness is considered to be represented by spherically shaped islands and standard values from Sipe's theory for the 'shape' ($s$ = 0.4) and the 'filling' ($f$ = 0.1) factors [18,19,53] are assumed. In this work, linearly polarised laser beams are used and according to Sipe's theory, LIPSS are formed where $\eta$-maps exhibits sharp features (maxima or minima). Simulation results are illustrated in Figs.5a,b,c at $\lambda_L$=2.6 μm for some representative values of the carrier densities $10^{19}$ cm$^{-3}$, $7\times10^{19}$ cm$^{-3}$ and $3\times10^{20}$ cm$^{-3}$. It is shown that sharp points of $\eta$ appear along the $K_y(= \lambda_L/\Lambda_y)$ axis (*black* dashed lines as shown in the insets) which indicates that LSFL are oriented *parallel* to the laser polarisation vector. Furthermore, as shown in the insets, the positions of the sharp features along $K_y$ yield the periodicities of the periodic structures to be equal to $\Lambda_y$=1.9 μm, 2.38 μm and 2.58 μm (i.e. through the expression $\Lambda_y = K_y/\lambda_L$), respectively. Therefore, Sipe's theory is capable to predict efficiently both the orientation and periodicities of LSFL∥. By contrast, for $N_e$=5×10$^{20}$ cm$^{-3}$, production of LSFL∥ disappear as no structures with sharp features along $K_y$ are predicted; by contrast, the $\eta$-map shows (Fig.5d) that another type of structures (termed LSFL⊥) is produced that are aligned *perpendicularly* to the laser polarisation. Simulation results illustrated in Fig.5d indicate that the electron density plays a significant role towards switching the orientation of the induced periodic structures from LSFL∥ to LSFL⊥. Similar conclusions were drawn at lower wavelengths, however, for remarkably higher values of $N_e$ [53]. The fact that slightly different electron densities result in a switch of orientation of the LIPSS by 90 degrees is a consequence of the different intensity values and absorption levels. This is an interesting outcome as it can be regarded as a saturation effect which can be employed in laser-based patterning techniques to produce either type of LIPSS in a controlled way by modulating the laser intensity (Fig.6).

Simulations were also conducted to derive a correlation between the electron densities, the laser intensities and the periodicities for LSFL∥ for four values of the laser wavelength $\lambda_L$=2.6 μm, 3 μm, 3.5 μm and 4 μm (Fig.6). In regard to the periodicity dependence on the electron density, results show that at relatively low densities the simulated predictions for $\Lambda_y$ scale as $\lambda_L/n$ (*red* line in Fig.6); by contrast, at larger $N_e$, the expression $\lambda_L/n$ cannot be used to calculate the periodicities of LSFL∥. Similar conclusions were also drawn at lower wavelengths [53]. Furthermore, ripple periodicities saturate to a value close to $\lambda_L$ at higher excitation levels and higher intensities.

In regard to the formation of LSFL⊥, besides Sipe's theory based on the interference of the incident light with the far-field of rough non-metallic or metallic surfaces [22], an alternative approach assuming SP excitation will be used to correlate the simulated electron density values and the predicted frequencies of the periodic structures (a similar scenario was considered [53] to explain the observed LSFL⊥ structures at very short pulses ~5fs [56]). This mechanism has been widely proposed for metals [1,13,14,24] and semiconductors [10,33], however, LSFL⊥ formation *based on the excitation of SP in dielectrics* represents a yet unexplored field. According to the SP-model, the calculated periodicity is provided by the expression $\Lambda = \lambda_L / Re\sqrt{\frac{\varepsilon}{\varepsilon+1}}$ [10,43] for irradiation in vacuum which is approximately correct for very small number of pulses (*NP*) [13]. Results in Fig.7 illustrate the computed values of $\Lambda$ as a function of $N_e$, both for the maximum electron densities calculated in this work (*red* dots) and for the range of values of $N_e$ that are necessary to satisfy the condition for SP excitation (i.e. $Re(\varepsilon)<-1$) [43,58]. Results are also used to derive the correlation between the intensity values that give rise to SP excitation and therefore LSFL⊥ periodic structures (Fig.7). Simulations were performed for four values of the laser wavelength $\lambda_L$=2.6 μm, 3 μm, 3.5 μm and 4 μm and results demonstrate a drop of $N_e$ and $I$ required to produce ripples at increasing laser wavelength.

The above paragraphs correlate the periodic features of LIPSS (periodicity and orientation) with the corrugation of the surface and the induced excited electron densities. Nevertheless, to provide a detailed description of LIPSS formation, a multiscale description of all physical processes that include energy absorption, electron excitation, transfer of heat from the electron to the lattice system, phase transitions and surface modification processes. More specifically, the following TTM (Eq.2) is used to describe the increase of the electron temperature after excitation as well as the relaxation process and lattice temperature change following energy transfer from the electron system to the lattice

$$\begin{aligned} C_e \frac{dT_e}{dt} &= \vec{\nabla}(k_e \vec{\nabla} T_e) - g(T_e - T_L) + S \\ C_L \frac{dT_L}{dt} &= \vec{\nabla}(k_L \vec{\nabla} T_L) + g(T_e - T_L) \end{aligned} \quad (2)$$

where $C_e$ and $C_L$ stand for the heat capacities of the electron and lattice subsystems, respectively while $T_e$ and $T_L$ are the temperatures of the two systems. On the other hand, $k_e$ ($k_L$) correspond to the electron (lattice) conductivity, $g$ is the electron-phonon coupling and $S$ is a source term that describes the average energies of the particles which are excited with the laser beam. A more detailed description of Eqs.2 is provided in the Supplementary Material. It is noted that the employment of Eqs.2 is based on the consideration that the laser pulse duration is long enough to assume an instantaneous thermalisation of the electron system through electron-electron scattering processes. By contrast, for very short pulses (<100 fs), a more thorough investigation is required to describe both the electron excitation, ultrafast dynamics and relaxation processes, however, this is beyond the scope of the present study.

To model the surface modification, it is assumed that the laser fluence is sufficiently high to result in a phase transition from solid to liquid phase and upon resolidification, a nonflat relief is induced on the surface of the material. Depending on the laser intensity, mass removal is also possible if the material is heated above a critical temperature (~ $T_L$>1.8 × $T_{boiling}$ for SiO$_2$ where $T_{boiling}$ = 2270 K [70]; the choice of the critical temperature is explained in the Supplementary Material). The movement of a material in the molten phase is given by the following Navier-Stokes equations (NSE) which describes the dynamics of an uncompressible fluid

$$\rho_0 \left( \frac{\partial \vec{u}}{\partial t} + \vec{u} \cdot \vec{\nabla} \vec{u} \right) = \vec{\nabla} \cdot \left( -P + \mu(\vec{\nabla}\vec{u}) + \mu(\vec{\nabla}\vec{u})^T \right) \quad (3)$$

where $\rho_0$ and $\mu$ stand for the density and viscosity of molten SiO$_2$, while $P$ and $\vec{u}$ are the pressure and velocity of the fluid. A more detailed description of the fluid dynamics module and numerical solution of NSE are provided in the Supplementary Material). In Eq.3, superscript $T$ denotes the transpose of the vector $\vec{\nabla}\vec{u}$ [10].

A multiple pulse irradiation scheme is required to derive the formation of periodic relief [10,11] through the solution of Eqs.1-3. More specifically, LIPSS are formed in the following steps (an analytical description of the numerical solution is provided in the Supplementary Material)

- the first pulse irradiates a flat surface and it is assumed to melt/ablate (depending on the laser intensity) part of the

material leading to the formation of a crater and humps at the edges on the surface of the heated zone due to mass displacement [10,11]. As the first pulse irradiates a flat surface with no corrugations, formation of periodic structures is not expected to occur [19]. It is noted that due to the axial symmetry of a Gaussian beam, for $NP=1$, Eqs.1-3 can be solved in 2D ($r$ and $Z$ where $r$ is the horizontal distance from the position of the maximum energy absorption while $Z$ stands for the vertical distance from the surface)[10],

- the second pulse irradiates the previously formed profile and therefore the spatial symmetry breaks; as a result, a 2D modelling can no longer be used. Based on the discussion in the beginning of the section, the coupling of the electric field of the incident beam with the induced surface scattered wave produces a nonuniform, periodic distribution of the absorbed energy (of periodicity and orientation determined by the density of excited carriers and the efficacy factor). More specifically, the $\eta$-map dictates the propagation direction of the spatial modulation of the deposited energy (Fig.5) with a $N_e$-dependent periodicity $\Lambda_y$ (Fig.6) [8,21]. In the present model, the induced spatial modulation of the absorbed energy is introduced as a sinusoidal function of periodicity $\Lambda_y$ in Eq.1,2 [13] (this approach allows modelling formation of LSFL∥; on the other hand, at higher excitation levels when LSFL⊥ structures are produced, as explained above, the periodicity can be determined either by the use of the efficacy factor map or the SP-model). The periodic variation of the absorbed energy leads to a periodic excited electron density distribution which is simulated with Eq.1. It is noted, though, that the computation of the amount of the absorbed energy at each position requires the evaluation of the energy deposition on a curved surface [10,13]. Therefore, appropriate computational schemes have been developed to compute the absorbed energy on each point of the curved surface [10]. The spatially modulated electron energy distribution is transferred to the lattice system (through Eq.2) and subsequently, upon phase transition (Eq.3), fluid transport and resolidification processes, LIPSS are formed.

- In an iterative fashion, each subsequent pulse irradiates a periodic pattern that has been created in the previous step; it is noted that the depth of the surface profile is increased with increasing $NP$ [9,13] while the periodicity of the induced pattern is always determined by the carrier density as mentioned in the previous section. The temporal separation between subsequent pulses is long enough to ensure that each pulse irradiates material in solid phase. These multiscale mechanisms have been simulated and observed in other irradiation conditions and at lower wavelengths in previous reports [10-14,18,22].

As an example, results for a multipulse simulated experiment are shown in Fig.8. The spatio-temporal evolution of the electron density following irradiation of fused silica with one pulse ($NP=1$) of wavelength $\lambda_L=2.6$ μm and a laser peak intensity $I=1.4\times10^{13}$ W/cm$^2$ is displayed in Fig.8a. Assuming that the electron density which leads to optical breakdown is equal to $N_e^{cr}=1.6561\times10^{20}$ cm$^{-3}$ for irradiation at this wavelength, a region for which $N_e > N_e^{cr}$ extends to ~490 nm (dashed line in Fig.8a). The inset in Fig.8a illustrates the temporal evolution of $N_e$ at 490 nm. The value of $N_e^{cr}$ has been usually attributed to the onset of damage [62] in spite of the fact that mass removal or melting occurs at longer times and after sufficiently high energy is transferred to the lattice system. By contrast, in this work, a thermal criterion is used (i.e. damage occurs when lattice temperature exceeds the melting point of the material) that constitutes a more precise estimation of the damage [11,12,49]. It is emphasised that the consideration of the peak intensity in the calculations was to compute the maximum depth of the heated region. Given the (spatially) Gaussian profile of the beam, it is evident that the maximum depth occurs where the absorbed energy is maximum. Figs.8b,c illustrates a top view (quadrant) of the surface profile for $\lambda_L=2.6$ μm and laser peak intensities of $1.06\times10^{13}$ W/cm$^2$ and $1.4\times10^{13}$ W/cm$^2$, respectively, for $NP=3$ and $NP=10$ respectively. Based on the above discussion, the maximum depth occurs at $X=Y=0$ (the position of the Gaussian intensity peak). The FWHM diameter of the laser spot is equal to 30μm. The two intensities correspond to energies that lead to the formation of LSFL∥ and LSFL⊥, respectively. The insets in Figs.8b,c illustrate the depth profile along the *white* dashed line which shows a pronounced periodic spatial profile. The predicted periodicities for the two patterns are 2.4 μm and 2.2 μm.

A fundamental question that rises about the calculation of the induced periodicities as a function of the irradiation dose is whether the models centred on (i) Sipe's or (ii) SP-based theories are valid with increasing $NP$. With respect to the former model, it is already known that one of the limitations of the efficacy factor-based theory is the neglect of the so-called 'feedback mechanism' which is very important to calculate the evolution of the periodicity of the induced periodic structures [19]. In the current work, firstly, simulations for $NP=2$ for laser peak intensities of $1.06\times10^{13}$ W/cm$^2$ were performed assuming the conventional Sipe's theory (i.e. $NP=1$ corresponds to the irradiation of a flat profile and therefore Sipe's theory cannot be used). The application of the multiscale model developed in this work and the use of Sipe's theory predict a maximum value of the carrier density equal to ~0.7×10$^{21}$ cm$^{-3}$ that yields a periodicity ~2.35 μm. It is assumed that similar values for 'shape' and 'filling' parameters ($s$ and $f$, respectively) can be used for $NP=3$ assuming a relatively similar corrugation and comparable carrier density; hence, a computed value equal to approximately ~2.4 μm is derived for $NP=3$ (Fig.8b). In all cases, LSFL∥ are produced (as shown in Figs.5a-c). On the other hand, as the depth of the surface pattern increases and its corrugation changes, a calculation of interpulse evolution of LSFL∥ periodicities with the employment of Sipe's theory becomes problematic.

By contrast, in Fig.8c, the periodic profile is illustrated for $NP=10$ following irradiation with $1.4\times10^{13}$ W/cm$^2$. Results in Fig.7a indicate that, for an almost flat profile, this laser intensity yields a carrier density ~$1.47\times10^{21}$ cm$^{-3}$ that leads to a SP periodicity equal to 2.55 μm. On the other hand, the computed value for the ripple periodicity for $NP=10$ is equal to ~2.2 μm. The latter value differs from the one computed through the expression $\Lambda = \lambda_L/Re\sqrt{\frac{\varepsilon}{\varepsilon+1}}$ which holds for nearly flat surfaces as enhanced corrugation has proven to yield a shift to the SP resonance to smaller values of $\Lambda$ at increasing

*NP*; this is also shown in previous studies in which lower laser wavelengths were used to fabricate LSFL⊥ in metals or semiconductors [9,10,13,23]. In contrast to electrodynamics simulations, mainly, based on Finite Difference Finite Domain Schemes (FDTD) used to correlate the induced LIPSS periodicities with a variable corrugation as a result of increase of the irradiation dose [9,22,48,71,72], an alternative and approximating methodology was employed, in this work, to relate the SP wavelength with the produced maximum depth of the corrugated profile [13,14] (i.e. which is linked with *NP*). The methodology was based on the spatial distribution of the electric field on a corrugated surface of particular periodicity and height and how continuity of the electromagnetic fields influences the features of the associated SP. The variation of the SP wavelengths (which also determines the LSFL⊥ periodicity) as a function of *NP* are shown in the Supplementary Material. Although, this technique provided theoretical predictions that agreed sufficiently well with experimental results [14], an FDTD-based analysis is mainly considered to represent a more accurate tool to describe the electrodynamical effects and evaluate periodicities of LSFL⊥ and LSFL∥.

The multiscale model used in this work showed the formation of LSFL∥ and LSFL⊥ while a transition from one type to another (considering the discussion about Fig.5), however, the types of structures that are produced for small number of *NP* (i.e. formation of LSFL∥) differs from results in previous combined theoretical simulations [22] and experimental observations [53,73] at *shorter* wavelengths. More specifically, previous studies showed that for small number of pulses and fluences, HSFL structures are formed on the surface of the material while LSFL∥ are induced at greater depths. The mixture of the two types of structures is also witnessed in diffraction experiments. An increase of the fluence or higher irradiation dose leads to a removal of the HSFL structures while LSFL∥ structures remain. Further irradiation leads to a transition from LSFL∥ to LSFL⊥. Certainly, a future revised version of the model including more precise electromagnetic simulations would allow to understand further LIPSS evolution with laser sources in the mid-IR spectral region.

It should be emphasised that, a more accurate and convincing conclusion about the validity of the aforementioned model and theoretical predictions to describe physical mechanisms in the mid-IR region for fused silica will be drawn if appropriately developed experimental protocols are introduced. To the best of our knowledge, there are no previous experimental reports that could provide a validation of the aforementioned model for $SiO_2$ in that spectral region. However, employment of Eq.1 (to compute electron excitation [49,50,52,62,70]), Sipe theory mechanisms (to evaluate LSFL periodicities [11,12,19,53,74]) and Eq.1-3 (to derive a multiscale description of the surface patterning mechanisms [11] with experimental validation [12,53,74]) showed that the theoretical model presented in this study is capable to explain efficiently physical processes at *shorter* wavelengths.

Moreover, a question that is raised with respect to the frequencies of the induced periodic structures is the impact of STE. The theoretical model presented in this work (and results shown in the Supplementary Material) shows, firstly, that there is a conspicuous influence of the STE on the carrier densities that, in turn, affect the LIPSS periodicities. Previous experimental results related to the investigation of femtosecond diffraction dynamics of LIPSS on fused silica at lower wavelengths [73], showed that STE are capable to influence the refractive index values and facilitate emergence of incubation effects. Therefore, similar experimental protocols are important for the comparison of results from the proposed STE-based model with experimental findings at longer wavelengths.

In conclusion, it is evident that the lack of experimental results might hinder the validation of the theoretical framework at *longer* wavelengths. An extension of the model used for semiconductors (i.e. Silicon [43]) to simulate damage thresholds was also recently validated by appropriate experimental protocols [38,43]. In regard to the limitations of the model, there are some yet unexplored issues that need to be addressed (i.e. investigation of behaviour in ablation conditions, consideration of formation of voids inside the material after repetitive irradiation, role of incubation effects [22], STE and defects [75], validity of the use of Eqs.1-2 for very short pulses where a more precise quantum mechanical approach is required to describe ultrafast dynamics [76], influence of ambipolar electron-hole plasma diffusion to damage thresholds [77], etc.) before a complete picture of the physical processes with femtosecond mid-IR laser pulses is attained. Nevertheless, the model is aimed to set the basis for a description of the multiscale processes that lead to surface modification in the mid-IR spectral region through the evaluation of the impact of the long pulses on various fundamental processes. On other hand, apart from the importance of elucidating the underlying mechanisms from a physical point of view, a deeper understanding of the response of the material will allow a systematic novel surface engineering with strong mid-IR fields for advanced industrial applications.

## Conclusions

To summarise, it is known that while an extensive research has been conducted towards elucidating laser-induced growth of damage for irradiation of $SiO_2$ with IR (or shorter) pulses, little is known about the effects of electron excitation with longer wavelength pulses. In this work, a detailed theoretical framework was presented for the first time that describes both the ultrafast dynamics and thermal response following irradiation of fused silica with ultrashort pulsed lasers in the mid-IR spectral region. The influence of nonlinearities in the refractive index, the ultrafast dynamics in a wide range of wavelengths and various intensities, as well as fundamentals of laser-based surface patterning, were investigated. There is no doubt that our theoretical approach requires validation and possibly further development before it can fully account for the physical processes taking place upon laser-material interaction in the mid-IR spectral region. Nevertheless, the predictions resulting from the above theoretical approach demonstrate that unravelling phenomena during such interaction can potentially set the basis for the development of new tools for a large range of mid-IR laser-based applications.

## Methods

**Computational method.** Full details of the multiscale model, numerical solution of the model and the parameters used are presented in the Supplementary Material.

**Acknowledgements**
The authors acknowledge financial support from *Nanoscience Foundries and Fine Analysis (NFFA)–Europe* H2020-INFRAIA-2014-2015 (under Grant agreement No 654360), *HELLAS-CH project* (MIS 5002735), implemented under the "Action for Strengthening Research and Innovation Infrastructures," funded by the Operational Programme "Competitiveness, Entrepreneurship and Innovation" and co-financed by Greece and the EU (European Regional Development Fund), COST Action *TUMIEE* (supported by COST-European Cooperation in Science and Technology). We would also like to acknowledge assistance of Anna Tsibidi in the preparation of Fig.1.

**Author contributions statement**
G.D.T. and E.S. conceived the idea, G.D.T performed the physical processes modelling, conducted the simulations and analysed the results. Both authors contributed to the preparation of the manuscript.

*Supplementary Material to* 'Ionisation processes and laser induced periodic surface structures in dielectrics with mid-infrared femtosecond laser pulses'

George D. Tsibidis[1,*], and Emmanuel Stratakis[1,2]

[1] Institute of Electronic Structure and Laser (IESL), Foundation for Research and Technology (FORTH), N. Plastira 100, Vassilika Vouton, 70013, Heraklion, Crete, Greece

[2] Department of Physics, University of Crete, 71003 Heraklion, Greece

Email: *tsibidis@iesl.forth.gr

## I. THEORETICAL MODEL

### A. Electron density rate equations

The absorption of light in transparent materials is a nonlinear process as a single photon does not have enough energy to excite electrons from the valence to the conduction band. Following nonlinear photo-ionisation (where both multiphoton and tunnelling ionisation can occur in different regimes depending on the laser intensity), impact ionisation and formation of STE lead to a variation of the electron densities. To calculate the laser absorption, the densities of the excited electron and the STE are computed by solving the following set of equations

$$\begin{aligned}\frac{dN_e}{dt} &= \frac{N_v - N_e}{N_v}\left(W_{PI}^{(1)} + N_e AI_r^{(1)}\right) + \frac{N_{STE}}{N_v}\left(W_{PI}^{(2)} + N_e AI_r^{(2)}\right) - \frac{N_e}{\tau_{tr}} \\ \frac{dN_{STE}}{dt} &= \frac{N_e}{\tau_{tr}} - \frac{N_{STE}}{N_v}\left(W_{PI}^{(2)} + N_e AI_r^{(1)}\right)\end{aligned} \quad (1)$$

where $N_V = 2.2 \times 10^{22}$ cm$^{-3}$ [49] corresponds to the atomic density of the unperturbed material while $N_e$ and $N_{STE}$ denote the free electron and STE electron densities, respectively. The last term in the first equation corresponds to free electron decay that is characterised by a time constant $\tau_{tr}$ ($\tau_{tr} \sim 150$fs in fused silica) that leads to a decrease of the electron density. By contrast, $W_{PI}^{(i)} + N_e AI_r^{(i)}$ ($i=1,2$) correspond to the combined photo-ionisation $W_{PI}^{(i)}$ and impact ionisation; $AI_r^{(i)}$ stands for the (usually termed) avalanche ionisation rate coefficients [49,60] of the impact ionisation term that depends on which bandgap the electrons need to surpass ($E_G^{(1)}$ (=9eV) or $E_G^{(2)}$ (=6eV)). The photoionisation rates are computed using the Keldysh formulation [60]

$$W_{PI}^{(i)} = \frac{2\omega}{9\pi}\left(\frac{m_r \omega_L}{\gamma_2 \hbar}\right)^{3/2} \Theta(\gamma, x) \exp\left[-\pi\langle x+1\rangle \frac{K(\gamma_2) - E(\gamma_2)}{E(\gamma_1)}\right] \quad (2)$$

where $\gamma = \omega_L \sqrt{m_r E_G^{(i)}}/(e|\vec{E}|)$ is the Keldysh parameter for the band-gap $E_G^{(i)}$ and is dependent on the electron charge $e$, the frequency $\omega_L$ and the field $|\vec{E}|$ of the laser beam, the electron reduced mass $m_r = 0.5 m_e$ ($m_e$ is the electron mass) and $\gamma_2 = \gamma/\sqrt{1+\gamma^2}$ and $\gamma_1 = \gamma_2/\gamma$. Furthermore, $\langle x+1 \rangle$ stands for the integer part of the number [42] $x+1$, where $x = 2E_G^{(i)} E(\gamma_1)/(\pi\gamma_2 \hbar \omega_L)$ while $K$ and $E$ are the complete elliptic integrals of the first and second kind, respectively. Also,

$$\Theta(\gamma, x) = \sqrt{\frac{\pi}{2K(\gamma_1)}} \sum_{N=0}^{\infty} \exp\left[-N\pi \frac{K(\gamma_2) - E(\gamma_2)}{E(\gamma_1)}\right] \times \Phi\left[\sqrt{\frac{\pi^2(2\langle x+1 \rangle - 2x + N)}{K(\gamma_1) E(\gamma_1)}}\right] \quad (3)$$

where $\Phi(z) = \int_0^z exp(y^2 - z^2) dy$. Finally, the coefficient $AI_r^{(i)}$ is given by the following expression

$$AI_r^{(i)} = \frac{e^2 \tau_c I}{\left[c\varepsilon_0 n m_r \left(\omega_L^2 (\tau_c)^2 + 1\right)(2 - m_r/m_e)\right]} \frac{1}{E_G^{(i)*}} \quad (4)$$

where $E_G^{(i)*} = E_G^{(i)} + e^2 |\vec{E}|^2 / (4 m_r \omega_L^2)$ is the effective band gap which takes into account the oscillation energy of the free electrons in the electric field, and the conservation of energy and the momentum during the collision between the free and bound electrons. Also, $c$ is the speed of light, $\varepsilon_0$ stands for the vacuum permittivity, $n$ is the refractive index of the material, while $I$ is the peak intensity of the laser beam, respectively, and $\tau_c$ is the electron collision time.

In regard to the value of $\tau_c$, there are many reports in which the collision time has been considered to be a constant or a varying parameter. More specifically, values of $\tau_c$ vary in a range between 0.1 fs and 10 fs [78] while in other studies values equal to 1.0 fs [79], 1.5 fs [80], 23.3 fs [54], 10 fs [81], 1.7 fs [82], 0.5 fs [22] have been reported. Thus, simulations with the above choices for $\tau_c$ yielded good agreement with experimental data.

On the other hand, there are other reports in which a more rigorous analysis was followed to compute the collision time: In Chimer et al [49], the contribution of electron-phonon, electron-electron, electron-neutral and electron-ion collision frequencies were used to compute $\tau_c$. In other reports [52,83,84], the electron density distribution function was also considered to calculate the electron collision frequency and the electron collision time. A remarkable impact of the value of the collision frequency on observable quantities such as the damage threshold has been demonstrated with simulations in other materials (i.e. Silicon [85]) which emphasized the need for a precise evaluation. Nevertheless, in the absence of experimental results in the mid-IR, the aim of the present work was firstly to provide a consistent multiscale modelling approach in which some approximations have been made rather than to consider a more precise evaluation of the collision frequency that could be the subject of revised model. Therefore, the collision frequency is taken constant in this work. In regard to the value chosen in this manuscript, the motivation was based on the work of Burakov et al [80] in which the damping term $\omega_L \tau_c = 3$ was used that resulted into a collision time equal to 1.5 fs (for $\lambda_L = 800$ nm). In the case of mid-IR, for laser wavelengths between 2 μm-4 μm, the collision time was calculated from the above expression yielding values for $\tau_c$ in the range [3.1fs, 6.4fs].

The dependence of the avalanche ionization rate coefficient $AI_r^{(i)}$ on the collision time to which carrier-carrier or carrier-ion scattering times (and dephasing oscillations) contribute [86,87] implies that a possible future revision of the model should include a term for impact ionization in which the collision time is a varying parameter; In relevant reports [86,87], an electron temperature dependent expression for $\tau_c$ was considered.

The photoionisation rates $W_{PI}^{(1)}$ and $W_{PI}^{(2)}$ include both multiphoton and tunneling ionisation processes that are provided from the following expressions

$$MPI^{(i)} = \frac{2\omega_L}{9\pi}\left(\frac{m_r \omega_L}{\hbar}\right)^{3/2} \Phi\left(\sqrt{2\left\langle \frac{E_G^{(i)}}{\hbar\omega_L}+1\right\rangle - 2\frac{E_G^{(i)}}{\hbar\omega_L}}\right) \times \exp\left(2\left\langle \frac{E_G^{(i)}}{\hbar\omega_L}+1\right\rangle\left(1 - \frac{1}{4\gamma^2}\right)\right)\left(\frac{1}{16\gamma^2}\right)^{\left\langle \frac{E_G^{(i)}}{\hbar\omega_L}+1\right\rangle} \quad (5)$$

$$TI^{(i)} = \frac{2E_G^{(i)}}{9\pi^2 \hbar}\left(\frac{m_r E_G^{(i)}}{\hbar^2}\right)^{3/2}\left(\frac{e\hbar|\vec{E}|}{\sqrt{m_r}\left(E_G^{(i)}\right)^{3/2}}\right)^{5/2} \exp\left(-\frac{\pi E_G^{(i)}}{2\hbar\omega_L \gamma}\left(1 - \frac{\gamma^2}{8}\right)\right) \quad (6)$$

The two ionisation processes become more efficient in different regimes; more specifically, tunneling ionisation becomes dominant for $\gamma \ll 1$ while multiphoton ionisation becomes more efficient for $\gamma \gg 1$ [60].

In contrast to the avalanche ionisation rate coefficient and impact ionisation, multiphoton and tunneling ionisation rates do not appear to be dependent on dephasing effects.

With respect to the intensity of the laser beam $I$, previous studies consider that the attenuation of the local laser intensity is determined by multiphoton ionisation and inverse bremsstrahlung (Free Carrier) absorption [22]. Nevertheless, the presence of STE states and the possibility of retransfer of carriers in these states back to the conduction band through multiphoton ionisation require modification of the spatial intensity distribution; therefore a revised 3D model (in Cartesian coordinates: $\vec{x} = (X, Y, Z)$) should include those contributions [11,49,61]

$$\frac{\partial I(t,\vec{x})}{\partial t} = -N_{ph}^{(1)} \hbar\omega_L \frac{N_v - N_e}{N_v} PI^{(1)} - \alpha(N_e) I(t,\vec{x}) - N_{ph}^{(2)} \hbar\omega_L \frac{N_{STE}}{N_v} PI^{(2)}$$

$$I(t,X,Y,Z=0) = (1-R(t,X,Y,Z=0))I_{peak} \exp\left(-4\log(2)\left(\frac{t-3\tau_p}{\tau_p}\right)^2\right) \exp\left(-\frac{X^2+Y^2}{(R_0)^2}\right)$$

(7)

where $N_{ph}^{(i)}$ corresponds to the minimum number of photons necessary to be absorbed by an electron that is in the valence band ($i$=1) or the band where the STE states reside ($i$=2) to overcome the relevant energy gap and reach the conduction band. $I_{peak}$ ($\equiv \frac{2J\sqrt{ln2}}{\sqrt{\pi}\tau_p}$) corresponds to the peak value of the laser intensity, $J$ stands for the laser fluence, $R(t,x,y,z=0)$ stands for the reflectivity of the material while $R_0$ is the irradiation spot radius ($R_0$=15μm in our simulations). The second equation of Eq.7 gives the spatial distribution of the intensity profile in Cartesian coordinates (and on the surface of the irradiated material).

It is also noted that, for the sake of simplicity, it is assumed that the beam shape does not change upon propagation; thus, only attenuation losses are taken into account and no Kerr or plasma distortion to the pulse phase were considered. A more accurate description of the laser energy distribution has been presented in other studies, by solving Maxwell's equations [22] or by including the shape change and Kerr effect to evaluate damage in the bulk [80]. However, we believe that a more accurate expression for the intensity profile would not lead to substantially different surface effects.

### B. Dielectric constant

The refractive index of fused silica when it is in an unexcited state is denoted by $n_0$ [88] (at $I$=0) ant it provided by the following expression

$$n_0 - 1 = \frac{0.6961663}{\lambda_L^2 - 0.0684043^2}\lambda_L^2 + \frac{0.4079426}{\lambda_L^2 - 0.1162414^2}\lambda_L^2 + \frac{0.8974794}{\lambda_L^2 - 9.896161^2}\lambda_L^2 \quad (8)$$

The expression $\varepsilon_{un} = n_0^2$ yields the dielectric constant of the unexcited material. Following irradiation of SiO$_2$ with femtosecond pulses, a temporally dependent density of excited carriers is produced that modify the dielectric constant of the material

$$\varepsilon' = 1 + (\varepsilon_{un} - 1)\left(1 - \frac{N_e}{N_v}\right) - \frac{e_c^2 N_e}{m_r m_e \varepsilon_0 \omega_L^2} \frac{1}{\left(1 + i\frac{1}{\omega_L \tau_c}\right)} \quad (9)$$

The reflectivity and free carrier absorption coefficients are given by the following expressions

$$\alpha_{FCA} = \frac{2\omega_L k}{c}$$

$$R = \frac{(1-n)^2 + k^2}{(1+n)^2 + k^2} \quad (10)$$

where $k$ is the extinction coefficient of the material. The real part of the refractive index $n$ is the sum of two terms, one ($n_0$) that corresponds the refractive index in an unexcited state and a second ($n_2 I$) that corresponds to the nonlinearities induced by Kerr effect (i.e. $n = n_0 + n_2 I$). $n_2$ is the Kerr coefficient. On the other hand, the dielectric constant is given by the following expressions if the Kerr effect is taken into account [89]

$$\varepsilon = \varepsilon' + \Delta\varepsilon_{kerr}$$

$$\Delta\varepsilon_{kerr} = 2n_0 n_2 I + (n_2 I)^2 \quad (11)$$

while $\varepsilon = (n+ik)^2$

### C. Electron and lattice heat balance

To describe the morphological properties due to laser irradiation of the material, it is important to explore, firstly, the relaxation process and heat transfer from the ionised material to the lattice system. Due to the metallic character of the excited material, a TTM model can describe the spatio-temporal dependence of the temperatures $T_e$ and $T_L$ of the electron and lattice subsystems, respectively [59]

$$\begin{aligned} C_e \frac{dT_e}{dt} &= \vec{\nabla}(k_e \vec{\nabla} T_e) - g(T_e - T_L) + S \\ C_L \frac{dT_L}{dt} &= \vec{\nabla}(k_L \vec{\nabla} T_L) + g(T_e - T_L) \end{aligned} \quad (12)$$

The source term has been modified properly to take into account all quantities that contribute to the total electron energy balance. Hence, the complete expression for source term $S(\vec{x},t)$ is given by [11,49,52]

$$\begin{aligned} S(\vec{x},t) &= \left(N_{ph}^{(1)} \hbar \omega_L - E_G^{(1)}\right) \frac{N_v - N_e}{N_v} PI^{(1)} - E_G^{(1)} AI_r^{(1)} N_e \frac{N_v - N_e}{N_v} \\ &+ \left(N_{ph}^{(2)} \hbar \omega_L - E_G^{(2)}\right) \frac{N_{STE}}{N_v} PI^{(2)} - E_G^{(2)} AI_r^{(2)} N_e \frac{N_{STE}}{N_v} \\ &+ \alpha(N_e) I(t,\vec{x}) \\ &- \frac{3}{2} k_B T_e \frac{N_e}{\tau_r} - \frac{3}{2} k_B T_e \frac{dN_e}{dt} \end{aligned} \quad (13)$$

The source term includes terms that are related to the dynamics of the total energy of the electron system. Processes that are taken into account, are the photoionisation of electrons (first and third terms), impact ionisation (second and fourth terms). It is noted that excitation both from VB (first and second terms) and STE (third and fourth) electrons are considered. Finally, the fifth term stands for the free electron absorption (that leads to transition to higher states in the CB) while the last two terms correspond to the energy balance due to electron density reduction (due to defect formation/trapping) and electron energy assuming carrier density variation [70]. Alternative expressions that incorporate average energies of electrons that are excited from STE states and exciton decay have been also proposed [52], however they have not been used in this work. The reason is that on the one hand, exciton decays are supposed to last very long (in the range of some hundreds of picoseconds [63,64]) compared to the timescales considered in the excitation process (and therefore their contribution is supposed to be insignificant); furthermore, the average energies of STE electrons contribution has been ignored in this study assuming it is not very significant compared that of the electron in CB. Nevertheless, a more precise investigation might be required in a future study by taking into account the role average energy of electrons in STE states in various laser conditions. We point out that a term that describes the divergence of the current of the carriers ($\vec{\nabla} \cdot \vec{J}$) has not been taken into account (it turns out that the consideration of particle transport of heat diffusion does not vary significantly quantities such as the damage threshold [85]). The temperature, electron density and temporal dependence of the thermophysical properties, $C_e$, $k_e$ are provided from well-established expressions derived from free electron gas based on the metallic character of the excited electron system [11,49]

$$\begin{aligned} E_F &= \frac{(hc)^2}{8 m_e c^2} \left(\frac{3}{\pi}\right)^{2/3} (N_e)^{2/3} \\ F(\varepsilon) &= \frac{8\sqrt{2}\pi (m_e)^{2/3}}{h^3} \sqrt{\varepsilon} \\ \mu(n_e, T_e) &= E_F \left[1 - \frac{\pi^2}{12}\left(\frac{k_B T_e}{E_F}\right)^2 + \frac{\pi^2}{80}\left(\frac{k_B T_e}{E_F}\right)^4\right] \\ \langle \varepsilon \rangle &= \frac{\int_0^\infty \exp\left(-\left((\varepsilon-\mu)/(k_B T_e)+1\right)\right) F(\varepsilon) \varepsilon d\varepsilon}{\int_0^\infty \exp\left(-\left((\varepsilon-\mu)/(k_B T_e)+1\right)\right) F(\varepsilon) d\varepsilon} \\ C_e(N_e, T_e) &= N_e \frac{\partial \langle \varepsilon \rangle}{\partial T_e} \\ k_e(N_e, T_e) &= \frac{1}{3}(u_e)^2 \tau_c C_e(N_e, T_e) \end{aligned} \quad (14)$$

On the other hand, $C_L$=1.6 J/(cm$^3$K) [49] while the coupling constant is estimated to be $g=g_0(N_e)^{2/3}$, where $g_0$=0.6×10$^{-1}$W/(mK) [49].

It is noted that the contribution of ambipolar diffusion of dense electron-hole plasma has not been included in the model. On the one hand, various studies have highlighted the role of ambipolar diffusion in the thermal response of the material and more specifically to studies related to damage threshold evaluation (see for example, studies by Danilov et al [77], and Derrien et al [90] on Silicon). On the other hand, in other studies [85], simulations did not show remarkable changes if the ambipolar diffusion (or transport) were ignored. It is evident though, that the influence might be dependent on the laser parameters (i.e. intensity, pulse duration, fluence and laser wavelength). Therefore this limitation of the present model could be the motivation for the development of a revised version.

Similarly, the temporal length of the pulse is also very important that can set limitations for the use of Eqs.1-14. One major parameter that influences the precision of the calculations is the pulse duration. The electron dynamics and relaxation processes considered in this study assumed an instantaneous thermalisation of the electron system through electron-electron scattering mechanisms. This is an important ingredient towards assuming that the electron system is in thermal equilibrium and $T_e$ is defined. By contrast, for very short pulses (for example, $\tau_p$<100 fs), this argument constitutes an overestimation and therefore alternative techniques to describe electron excitation and relaxation processes are required (see for example Ref. [24] and references therein). Alternative modelling approaches should be used in that case such as quantum mechanics-based approaches to account for the thermalisation of the electron system and dephasing effects. Similarly, appropriate corrections should be made if pulse durations smaller than the trapping time are used.

### D. Fluid dynamics

To model a surface modification following irradiation with mid-IR fs laser pulses, it is assumed that the laser conditions are sufficiently high to result in a phase transition from solid to liquid phase and upon resolidification a periodic relief is induced on the surface of the material based on the series of processes described in the main manuscript. Depending on the laser intensity, mass removal is also possible if the material is heated above a critical temperature (~ $T_L$>1.8 × $T_{boiling}$~4000 for SiO$_2$ where $T_{boiling}$ = 2270 K [70]). The choice of the critical temperature is based on arguments made in previous reports [10,14,70,91]. More specifically, a solid material subjected to ultrashort pulsed laser heating at sufficiently high fluences undergoes a phase transition to a superheated liquid whose temperature reaches values ~0.90 × $T_{critical}$ ($T_{critical}$ stands for the thermodynamic critical temperature [91]). In this work, the proposed scenario, of modeling material removal is based on a combination of evaporation of material volumes that exceed upon irradiation lattice temperatures close to ~0.90 × $T_{critical}$ and evaporation due to dynamics of Knudsen layer (adjacent to the liquid-vapor interface [22,25,26]). According to the discussion in Ref. [70], for many materials, a typical value of $T_{critical}$ is 1-2 times higher than the boiling temperature which is $T_{boiling}$ = 2270 K for fused silica [70], hence $T_{critical}$~2 × $T_{boiling}$ (assuming the smallest value for the attained value of $T_{critical}$). Based on this assumption, the minimum threshold value which should be used is equal to 2 × 0.9 × $T_{boiling}$ = 1.8 × $T_{boiling}$. Alternative scenarios for the estimation of the minimum lattice temperature that leads to mass removal could involve a lower temperature, the boiling temperature. It is evident that appropriate experimental setups could assist in a more precise evaluation, however, this is beyond the scope of the present study.

The movement of a material in the molten phase ($T_{melting}$=1988 K [12,92]) is given by the following Navier-Stokes equations (NSE) which describes the dynamics of an uncompressible fluid

$$\rho_0 \left( \frac{\partial \vec{u}}{\partial t} + \vec{u} \cdot \vec{\nabla} \vec{u} \right) = \vec{\nabla} \cdot \left( -P + \mu(\vec{\nabla}\vec{u}) + \mu(\vec{\nabla}\vec{u})^T \right) \tag{15}$$

where $\rho_0$ and $\mu$ stand for the density and viscosity of molten SiO$_2$, while $P$ and $\vec{u}$ are the pressure and velocity of the fluid. The fluid is considered to be an incompressible fluid (i.e. $\vec{\nabla} \cdot \vec{u} = 0$).

In regard to the pressure, there are two terms that require special treatment:
- the **recoil pressure** which is related to the lattice temperature of the surface of the material through the equation [93,94]

$$P_r = 0.54 P_0 \exp\left( L_v \frac{T_L^S - T_{boiling}}{R T_L^S T_{boiling}} \right) \tag{16}$$

where $P_0$ is the atmospheric pressure (i.e. equal to 10$^5$ Pa [95]), $L_v$ is the latent heat of evaporation of the liquid ($L_v$ =7 kJ/gr [95]), $R$ is the universal gas constant, and $T_L^s$ corresponds to the surface temperature. When vapour is ejected, it creates a back (recoil) pressure on the liquid free surface which in turn pushes the melt away in the radial direction[10] which results into a depression of the surface. Furthermore, given the spatially modulated energy deposition on the material, a gradient of the lattice temperature is produced which is, in turn, transferred into the fluid and therefore a capillary fluid convection is produced.

- A precise estimate of the molten material behaviour requires a contribution from the **surface tension related pressure**, $P_\sigma$, which is influenced by the surface curvature and is expressed as $P_\sigma = K\sigma$, where $K$ is the free surface curvature and $\sigma=0.310$ N/m [96] is the surface tension. The calculation of the pressure associated to the surface tension requires the computation of the temporal evolution of the principal radii of surface curvature $R_1$ and $R_2$ that correspond to the convex and concave contribution, respectively [97]. Hence the total curvature is computed from the expression $K=(1/R_1 + 1/R_2)$. A positive radius of the melt surface curvature corresponds to the scenario where the centre of the curvature is on the side of the melt relative to the melt surface.

Pressure equilibrium on the material surface implies that the pressure $P$ in Eq.15 should outweigh the accumulative effect of $P_r + P_\sigma$. The thermocapillary boundary conditions imposed at the liquid free surface are the following

$$\frac{\partial u}{\partial Z} = -\sigma/\mu \frac{\partial T_L}{\partial X} \quad \text{and} \quad \frac{\partial v}{\partial Z} = -\sigma/\mu \frac{\partial T_L}{\partial Y} \tag{17}$$

where $(u,v,w)$ are the components of $\vec{u}$ in Cartesian coordinates. Values for the thermophysical parameters that are used in the simulations are: $\rho_0 = 2.2$ gr/cm$^3$ [95], $\mu$ results from a fitting procedure (Ref.[96,98]) and $\sigma=0.310$ N/m [95]. The melting point of fused silica 1988K is taken as the threshold for a phase transition from solid to liquid while the same isothermal is considered as the same criterion for resolidification.

The hydrodynamic equations are solved in regions that contain either solid or molten material. To include the 'hydrodynamic' effect of the solid domain, material in the solid phase is modelled as an extremely viscous liquid ($\mu_{solid}=10^5 \mu$), which results into velocity fields that are infinitesimally small. An apparent viscosity is then defined with a smooth switch/step function

$$\delta(T_L - T_{melting}) = \frac{1}{\sqrt{2\pi}\Delta} e^{-\left[\frac{(T_L - T_{melting})^2}{2\Delta^2}\right]} \tag{18}$$

where $\Delta$ is in the range of 10-100$^0$K depending on the temperature gradient [10,70].

### E. Numerical scheme

To solve the set of the above equations, a scheme based on finite difference method is used. A common approach followed to solve similar problems is the employment of a staggered grid finite difference method which is found to be effective in suppressing numerical oscillations. Unlike the conventional finite difference method, temperatures ($T_c$ and $T_L$), carrier densities ($N_e$), pressure ($P$) are computed at the centre of each element while time derivatives of the displacements and first-order spatial derivative terms are evaluated at locations midway between consecutive grid points. For time-dependent flows, a common technique to solve the Navier-Stokes equations is the projection method and the velocity and pressure fields are calculated on a staggered grid using fully implicit formulations [99,100]. On the other hand, the horizontal and vertical velocities are defined in the centres of the horizontal and vertical cells faces, respectively (for a more detailed analysis of the numerical simulation conditions and the methodology towards the description of fluid dynamics, see Refs. [10,12-14,18,24,101,102]).

During the ultrashort period of laser heating, heat loss from the upper surface of target is assumed to be negligible. As a result, a zero heat flux boundary condition is set for the carrier and lattice systems.

- For irradiation with one pulse (*NP=1*), Eqs.1-17 are solved assuming heating of a flat profile; a 2D numerical approach is followed by taking into account the axial symmetry of the problem.

- For *NP>2*, the symmetry breaks and a 2D solution is no longer valid. In that case, a 3D numerical framework is developed (a finite difference methods is used again [10]). The incident beam is no longer perpendicular to the modified profile and therefore the surface geometry influences the spatial distribution of the deposited laser energy. Hence, appropriate modification to the numerical scheme is required to compute energy absorption (for example Eq.7 needs to be corrected). Typical Fresnel equations are used to describe the reflection and transmission of the incident light. Due to multiple reflection and light entrapment, the absorption of the laser beam is modified [103]. The calculation of the pressure associated to the surface tension requires the computation of the temporal evolution of the principal radii of surface curvature $R_1$ and $R_2$ that correspond to the convex and concave contribution, respectively [97]. Hence the total curvature is computed from the expression $K=(1/R_1 +1/R_2)$ [10].

In regard to the material removal simulation, in each time step, lattice and carrier temperatures are computed and if lattice temperature reaches $\sim T_L>1.8T_{boiling}$, mass removal through evaporation is assumed. In that case, the associated nodes on the mesh are eliminated and new boundary conditions of the aforementioned form on the new surface are enforced. In order to preserve the smoothness of the surface that has been removed and allow an accurate and non-fluctuating value of the computed curvature and surface tension pressure, a fitting methodology is pursued [10].

## F. Impact of fluence, pulse duration, laser wavelength on excitation levels.

To evaluate the role of STE states, the electron density and its variation has been calculated as a function of the laser fluence, the pulse duration and the laser wavelength (Figs.S1-S4) for single shot simulations. It is evident that the variation of the electron densities if STE states are formed (Eq.1 in the main manuscript) increases as the higher fluence increases. On the other hand, although simulation results (Figs.S1-S4) indicate that the role of STE can be ignored for small fluences, it is noted that experimental observations have shown that the presence of defects and incubation effects influence significantly the damage threshold of the irradiated material [61,75,104] for multipulse irradiation. Moreover, it is noted that in multiple shot experiments, the change of surface morphology (formation of crests and wells) vary the energy absorption, excited carrier densities and thermal response of the material and therefore a thorough investigation of the impact of repetitive irradiation (including the impact of STE) is required.

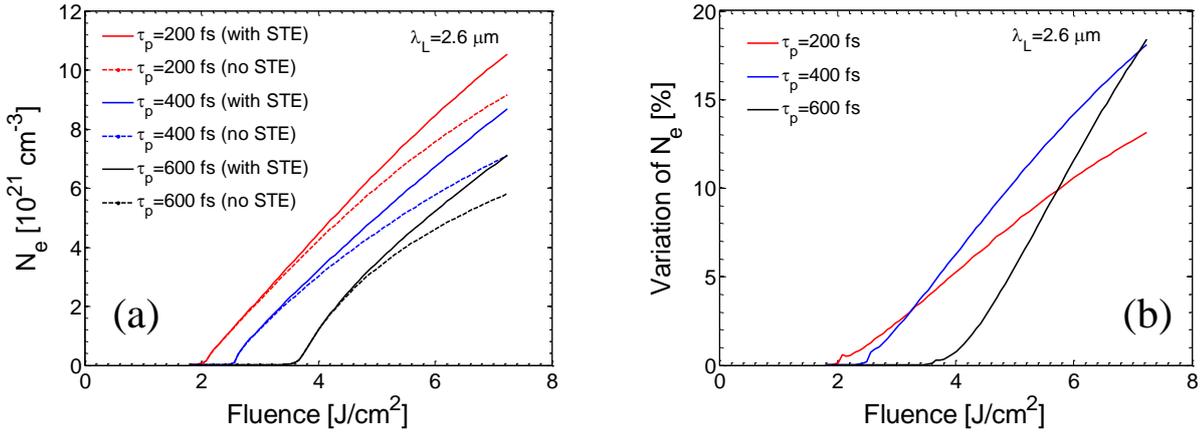

**Figure S1.** Electron densities (a) and percentage variation (b) with and without STE as a function of fluence for four values of the pulse duration ($\tau_p$=200 fs, 400 fs, 600 fs). Results are shown for $\lambda_L$=2.6 μm.

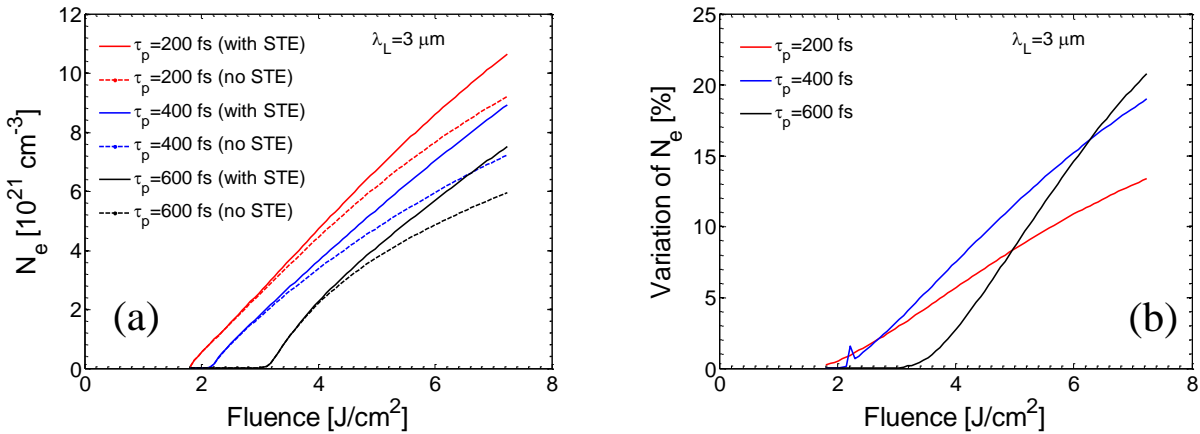

**Figure S2.** Electron densities (a) and percentage variation (b) with and without STE as a function of fluence for four values of the pulse duration ($\tau_p$=200 fs, 400 fs, 600fs). Results are shown for $\lambda_L$=3 μm.

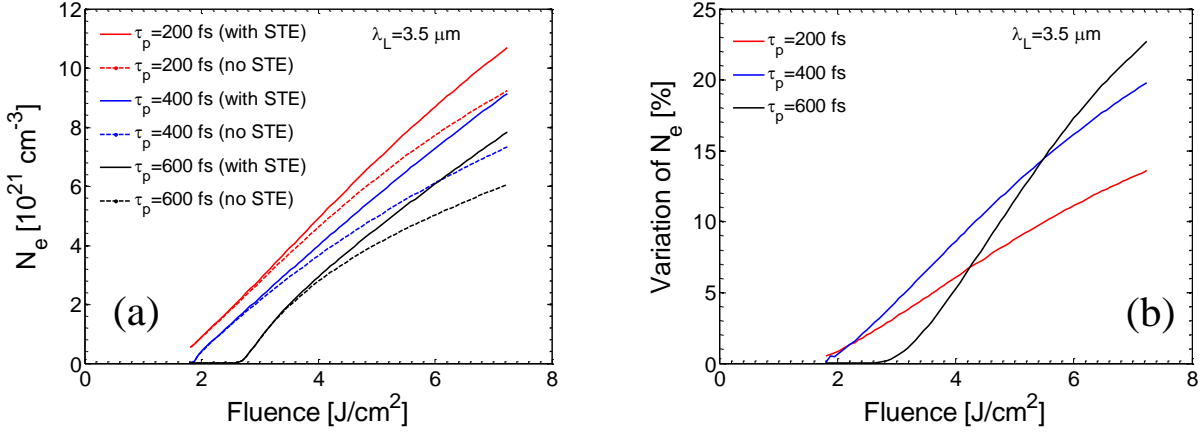

**Figure S3.** Electron densities (a) and percentage variation (b) with and without STE as a function of fluence for four values of the pulse duration ($\tau_p$=200 fs, 400 fs, 600fs). Results are shown for $\lambda_L$=3.5 μm.

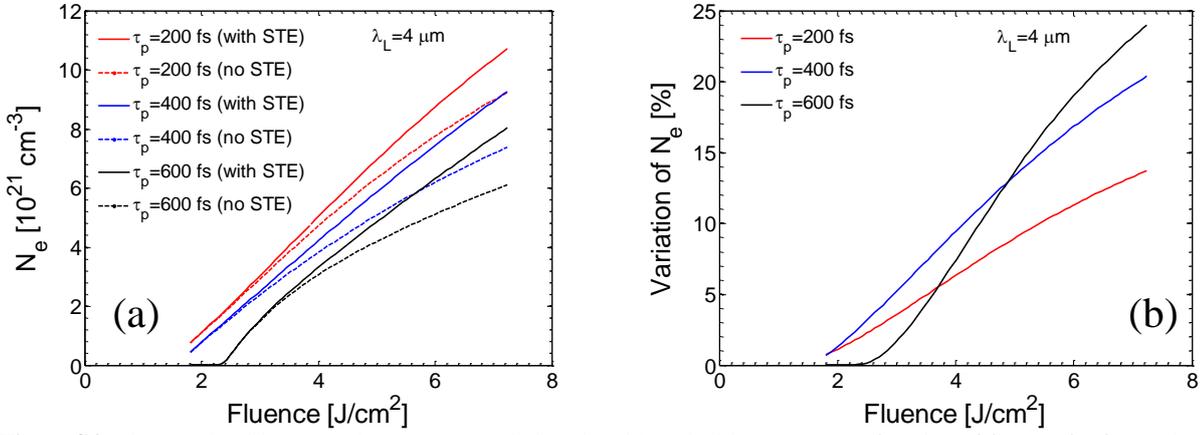

**Figure S4.** Electron densities (a) and percentage variation (b) with and without STE as a function of fluence for four values of the pulse duration ($\tau_p$=200 fs, 400 fs, 600fs). Results are shown for $\lambda_L$=4 μm.

## G. SP wavelength *vs*. NP.

Simulations have been performed to compute the variation of the SP wavelength with increasing irradiation dose (*NP*). To compute the SP wavelength, a multiscale approach has been followed to take into account the carrier density for the profile induced as a result of the surface profile after irradiation with the *NP*[th]-1 pulse. In Fig.S5, results are shown for $\lambda_L$=2.6 μm for SP wavelength as a function of *NP*.

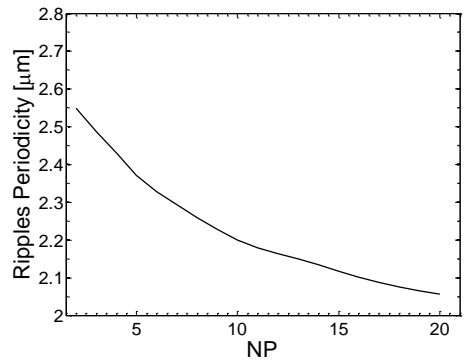

**Figure S5.** SP wavelength *vs*. NP. Results are shown for $\lambda_L$=2.6 μm (laser peak intensity $I$=1.4×10$^{13}$ W/cm$^2$).